\documentclass[11pt,english]{article}
\usepackage[T1]{fontenc}
\usepackage[latin9]{inputenc}
\usepackage{geometry}
\usepackage{hyperref}
\usepackage{cite}

\usepackage[usenames,dvipsnames]{color}
\usepackage{graphicx}
\usepackage{amsmath}
\usepackage{amsfonts}
\usepackage{amssymb}
\usepackage{dsfont}
\usepackage{dcolumn}
\usepackage{bm}
\usepackage{slashed}
\usepackage{pstricks}
\usepackage{slashed}
\usepackage{multirow}
\usepackage{array}
\usepackage{rotating}
\usepackage{xcolor}
\usepackage{epstopdf}
\usepackage{fancybox}
\usepackage{fancyvrb}
\usepackage[normalem]{ulem}
\usepackage{color}
\usepackage{mathdots}
\usepackage{mathrsfs}
\usepackage{multirow}
\usepackage{esint}
\allowdisplaybreaks

\newcolumntype{x}[1]{
{\centering}p{#1}}%

\def \d{{\rm d}}

\newcommand{\GeV}      {~\mathrm{GeV}}
\newcommand{\TeV}      {~\mathrm{TeV}}

\def \cha{\widetilde{\chi}^{\pm}_1}

\newcommand{\beqn}{\begin{eqnarray}}
\newcommand{\eeqn}{\end{eqnarray}}
\newcommand{\be}{\begin{equation}}
\newcommand{\ee}{\end{equation}}
\newcommand{\non}{\nonumber \\}

\newcommand{\mathsym}[1]{{}}

\def \cha{\tilde{\chi}^{\pm}_1}

\def \no{\tilde{\chi}^{0}}
\def \na{\tilde{\chi}^{0}_1}
\def \nsta{\tilde{\chi}^{\rm St}_1}

\def \nstb{\tilde{\chi}^{\rm St}_2}

\def \n34{\tilde{\chi}^{0}_{3,4}}
\def \g{\tilde{g}}

\def \ta{\tilde{t}_1}

\def \ba{\tilde{b}_1}

\def \sta{\tilde{\tau}_1}

\def \smr{\tilde{\mu}_R}
\def \ser{\tilde{e}_R}

\def\mhf{m_{1/2}}

\def\met100{\slashed{E}_T\geq 100 \GeV}
\def\sm{standard model~}

\newcommand{\st}{Stueckelberg~}
\def\Lag{{\mathcal{L}}}
\newcommand{\gappeq}{\mathrel{\rlap {\raise.5ex\hbox{$>$}}
{\lower.5ex\hbox{$\sim$}}}}
\newcommand{\lappeq}{\mathrel{\rlap{\raise.5ex\hbox{$<$}}
{\lower.5ex\hbox{$\sim$}}}}

\newcommand{\coup[1]}{\frac{1}{2}g_CQ_C^{#1}}

\def \gev{\text{~GeV}}
\def\zp{Z^{\prime}}

\def\met{\slashed{E}_{T}}

\begin{document}

\title{\textbf{
Cosmic Coincidence
and Asymmetric Dark Matter
in a Stueckelberg Extension}}
\author{Wan-Zhe~Feng\footnote{Email: vicf@neu.edu},
~Pran~Nath\footnote{Email: nath@neu.edu},
and Gregory~Peim\footnote{Email: peim.g@husky.neu.edu}
\\\textit{Department of Physics, Northeastern University, Boston, MA 02115, USA}}
\date{}
\maketitle

%%%%%%%%%%%%%%%%%%%%%%%%
%%%%%     abstract
%%%%%%%%%%%%%%%%%%%%%%%%

\begin{abstract}
We discuss the possibility of the cosmic coincidence  generating the ratio of baryon asymmetry to dark matter in a Stueckelberg $U(1)$ extension of the standard model and of the minimal supersymmetric standard model. For the $U(1)$ we choose $L_{\mu}-L_{\tau}$ which is anomaly free and can be gauged. The dark matter candidate arising from this extension  is a singlet of the standard model gauge group but is charged under $L_{\mu}-L_{\tau}$. Solutions to the Boltzmann equations for relics in the presence of asymmetric dark matter are discussed. It is shown that the ratio of the baryon asymmetry to dark matter consistent with the current WMAP data, i.e., the cosmic coincidence, can be successfully explained in this model with the depletion of the symmetric component of dark matter from  resonant annihilation via the Stueckelberg gauge boson. For the extended MSSM model it is shown that one has a two component dark matter picture with asymmetric dark matter being the  dominant component and the neutralino being the subdominant  component (i.e., with relic density  a small fraction of the WMAP cold dark matter value). Remarkably, the subdominant component can be detected in direct detection experiments  such as SuperCDMS and  XENON-100. Further, it is shown that the class of Stueckelberg models with a gauged $L_{\mu}-L_{\tau}$ will produce a dramatic signature at a muon collider with the $\sigma(\mu^+\mu^-\to \mu^+\mu^-,\tau^+\tau^-)$ showing a detectable $Z'$ resonance while $\sigma(\mu^+\mu^-\to e^+e^-)$ is devoid of this resonance. Within the above frameworks we discuss several broad classes of models both above and below the electroweak phase transition temperature. Asymmetric dark matter arising from a  $U(1)_{B-L}$ Stueckelberg extension  is also briefly discussed. Finally, in the models we propose the asymmetric dark matter does not oscillate and there is no danger of it being washed out from oscillations.
\end{abstract}

%%%%%%%%%%%%%%%%%%%%%%%%
%%%%%     introduction
%%%%%%%%%%%%%%%%%%%%%%%%

\newpage

\section{Introduction}
One of the outstanding puzzles in particle physics and cosmology relates to the so
called cosmic coincidence, i.e., the apparent closeness of the
amount of baryon asymmetry to the amount of dark matter in the Universe.
Thus the WMAP-7 result, with
{\sc RECFAST}~version~{\sc 1.5} to calculate the recombination history~\cite{wmap},
gives the baryonic relic density to be $100\Omega_{\rm B} h_0^2=2.255\pm 0.054$
and the dark matter relic density to be $\Omega_{\rm DM}h_0^2=0.1126\pm0.0036$, which
leads to
\beqn
\frac{\Omega_{\rm DM}h_0^2}{ \Omega_{\rm B}h_0^2} = 4.99\pm 0.20\,.
\label{ratio}
\eeqn
The closeness of $\Omega_{\rm DM}h_0^2$ and $\Omega_{\rm B}h_0^2$ points  to the possibility
that the baryonic matter and dark matter may have a common origin;  a possibility that
has been noted for some time~\cite{adm1}.
 In this manuscript we
analyze this issue  in the framework of a  \st $U(1)$ extension of the
\sm (SM) as well as a \st $U(1)$ extension of the minimal
supersymmetric standard
model~(MSSM)~\cite{kn12,FLNPRL,kctc,st2,st3,Feldman:2010wy,Liu:2011di}.
There are two main constraints  in building models  with asymmetric dark matter~(AsyDM).
First, we  need a mechanism for transferring a $B-L$
asymmetry produced in the early universe to dark matter.
Second, we must have a mechanism for depleting the symmetric component of dark matter
generated via thermal processes.
\\

The above issues have been discussed in the literature in  a variety of works
~\cite{adm2,Dreiner:1992vm,admBL,imy,gsv,admR, admS, admX, admE, admO}
(for a review see~\cite{Davoudiasl:2012uw}).
The models based on the \st extensions we discuss  in this work are different from the ones considered
previously both in terms of the mechanism for depletion of the symmetric component of dark matter
as well as regarding the implications for dark matter and signatures at colliders. Specifically, we consider a $U(1)_X$
extension of the \sm gauge group which is anomaly free. Further, we consider dark matter candidates
which will carry lepton number but not a baryon number, and are singlets of the \sm  gauge group.
Now in the leptonic sector it is known~\cite{He:1991qd} that for the \sm case
we may choose one of the linear combinations $L_e -L_{\mu}$, $L_{\mu}-L_{\tau}$,
$L_e - L_{\tau}$  to be anomaly free and can be gauged.
The gauged $L_e -L_{\mu}$ has been discussed previously in the context of PAMELA positron excess
 and multi-component
dark matter~\cite{Feldman:2010wy} and $L_{\mu}-L_{\tau}$ in the context of muon anomalous moment~\cite{Baek:2001kca}
and in the context the PAMELA positron excess~\cite{Baek:2008nz}. Here we consider a gauged
$L_{\mu}-L_{\tau}$ in the discussion of asymmetric dark matter as this choice is the more appropriate
one for the analysis here.
Specifically, we will consider a
$U(1)_{X}$, $X= L_{\mu}-L_{\tau}$ \st extension   of the  \sm as well as of the MSSM.
As is well-known, the MSSM
supplemented by supergravity soft breaking gives the neutralino as the lowest supersymmetric
particle and with $R$ parity a candidate for dark matter. Thus for the AsyDM to work in the MSSM extensions it is necessary to
have the neutralino as a subdominant component. This issue will be addressed as well as the
question if such a subdominant component may still be detectable in experiments for the direct detection
of dark matter. It is found that the \st  models with a gauged $L_{\mu}-L_{\tau}$ can produce
dramatic signatures at a muon collider leading to a detectable $\zp$ resonance in the $\mu^+\mu^-\to \mu^+\mu^-,
\tau^+\tau^-$ cross section while the $\mu^+\mu^-\to e^+e^-$ cross section exhibits
no such resonance.
Finally we consider the possibility of an asymmetric dark matter in the \st extension of $B-L$.
\\

An important issue regarding asymmetric dark matter concerns the possibility that such matter
can undergo oscillations~\cite{admO}. Thus, for example, consider a model which allows for a
Majorana mass term $\mathcal{L}= - m_M XX+h.c.$, where $X$ is the dark particle.
The presence of such a term along with other mass terms allows for the oscillation of $X$ to
its anti-particle $\bar X$. Detailed analysis show that in this circumstance
over the age of the Universe the asymmetric dark matter would produce a symmetric
component which would lead to pair annihilation.  Such processes could completely
wipe out the asymmetric dark matter generated in the early universe, and could render
such models largely invalid~\cite{admO}.
In the models we consider  mass terms
that can generate oscillations are forbidden because of gauge invariance, either
$U(1)_{L_\mu - L_\tau}$ or $U(1)_{B - L}$.   Thus  dark matter oscillations are
absent in the class of models we consider.
However, we wish to add a further explanation here. In general
there are two alternative possibilities for the \st
mechanism to arise. One possibility is that it is a low energy remnant of a Higgs  mechanism which
contains a scalar field $S$ that would in general allow a term of the type $XXS^n$ and thus generate
a Majorana mass term for $X$ as a consequence of VEV formation of the scalar field $S$.
This would lead to dark  matter oscillations once again.  The second
possibility, which is the view point adopted in this work, is that the \st mechanism arises
from the Green-Schwarz term in string theory (see, e.g.,~\cite{Ghilencea:2002da}).
In this case there is no fundamental Higgs field
which develops a VEV and thus Majorana mass term would not be generated and there would
be no dark matter oscillation.
\\

We will discuss six broad classes of models. Three of these will be anchored in extensions
of the standard model, one in extension of the two Higgs doublet model, and two in
extensions of the minimal supersymmetric standard model.
We will consider cases where the asymmetry transfer interaction may lie above or below the
electroweak phase transition scale, i.e., the scale where the Higgs boson gets its VEV.
In the supersymmetric case we will consider the case where all of the sparticles are in the
thermal bath at  temperatures where the asymmetry transfer takes place as well as the
case where the first two generations of squarks are heavy and are Boltzmann suppressed
in the thermal bath.
\\

The outline of the rest of the paper is as follows:
In Section~\ref{Sec:CgDMM} we give a brief introduction to  cosmic coincidence as well as asymmetric
dark matter and the technique for the computation of
the ratio of dark matter density to the baryonic matter density
in the Universe and of the dark matter mass.
Here we describe six broad classes of models which we will discuss in detail later.
In Section~\ref{Sec:CPSM} we carry out an explicit computation of
these quantities in extensions of the standard model and of the two Higgs doublet model.
In this section we also consider the case with inclusion of right-handed neutrinos.
In Section~\ref{Sec:CPMSSM} the analysis is redone for models in extensions of MSSM.
In Appendix~\ref{master} we give a master formula for the computation of the asymmetric dark matter
mass which is valid for temperatures above the electroweak phase transition scale. Here
we show that the results of models discussed in previous sections can be deduced as limiting cases.
In Section~\ref{Sec:ASY_SM} we consider an explicit \st extension of the \sm which generates asymmetric
dark matter. We discuss
solutions to the Boltzmann equations for relics in the presence of asymmetric dark matter,
and show that the symmetric component of
dark matter can be depleted from resonant annihilation via the $\zp$ pole. In Section~\ref{Sec:ASY_MSSM} we give a
\st extension of MSSM. Here a similar resonant annihilation of the symmetric component of dark
matter is valid.  There are several additional particles that arise in this case which include an extra
scalar particle (the $\rho$) from the \st sector. The decay width of this particle is computed in Appendix~\ref{App:rhoDR}
and it is shown that it decays rapidly and is removed from the relativistic plasma.
There are also additional neutralinos which we assume lie above the lightest MSSM neutralino
and thus the lightest MSSM neutralino continues to be the lightest supersymmetric particle~(LSP).
In Section~\ref{Sec:Ex_De} we show that the MSSM neutralino is
a subdominant component and thus does not interfere with the AsyDM mechanism.
It is also shown here that the subdominant component can
produce a spin-independent cross section which lies within reach of future experiments for the
direct detection of dark matter. In Section~\ref{Sec:Ex_Co} we discuss the signatures of the models at a muon collider.
A $(B-L)$ \st extension is discussed in Section~\ref{Sec:B-L} which also produces asymmetric dark matter.
Conclusions are given in Section~\ref{Sec:Con}.
In Appendix~\ref{App:Z'}, we compute the $\mu^+\mu^-\to e^+e^-$ at the loop level
via $\zp-\gamma$, $\zp-Z$ exchange and show that the corresponding production cross
section is too small to be discernible.

\section{Cosmic coincidence and  asymmetric dark matter} \label{Sec:CgDMM}
In the analysis here we assume that a $B-L$ asymmetry
has been generated in the early universe. We do not speculate on how this asymmetry comes
about as it could be by any number of different processes such as  baryogenesis or
leptogenesis~\cite{Buchmuller:2005eh}. Thus, for example, baryon asymmetry, specifically a non-vanishing $B-L$,
can arise in the early universe by decay of super-heavy particles in some grand unified
models~\cite{Enomoto:2011py,bm} consistent with experimental proton decay limits~\cite{Nath:2006ut}.
A $(B-L)$ asymmetry of this type will not be washed out by
sphaleron processes which preserve $B-L$.  Such an asymmetry is then transferred to
the dark sector at high temperatures via an interaction
of the form~\cite{admBL}
\begin{equation}
\mathcal{L}_{{\rm asy}}=\frac{1}{M_{{\rm asy}}^{n}}\mathcal{O}_{{\rm DM}}\mathcal{O}_{{\rm asy}}\,,
\label{ASYINT}
\end{equation}
where $M_{{\rm asy}}$ is the scale of this interaction,\footnote{
In the radiation-dominated era, the Hubble expansion rate is given by
$H\sim T^2 / M_{\rm Pl}$, where $M_{\rm Pl}=2.435 \times 10^{18}\GeV$ is the reduced Plank mass.
For an interaction suppressed by a factor $1/M^n_{\rm asy}$, the interaction rate at temperature $T$ is
$\Gamma(T)\sim T^{2n+1}/ M^{2n}_{\rm asy}$. Thus, the interaction will decouple if $\Gamma < H$, i.e., when
\begin{equation}
M^{2n}_{\rm asy} > M_{\rm Pl} T^{2n-1}\,. \label{MST}
\end{equation}
} and $\mathcal{O}_{{\rm asy}}$ is an operator constructed
from SM/MSSM fields which carries a non-vanishing $B-L$ quantum
number while $\mathcal{O}_{{\rm DM}}$ carries the opposite $B-L$
quantum number. This interaction would decouple at some temperature
greater than the dark matter mass.  As the Universe cools,
the dark matter asymmetry freezes on order of the baryon asymmetry, which explains the observed relation
between baryon and dark matter densities.
\\

At the temperature where Eq.~(\ref{MST}) is operational, and using the fact that the chemical potential
of particles and anti-particles are different,
the asymmetry in the particle and antiparticle number densities is given by
\beqn
n_i-\bar{n}_i = \frac{g_i}{2\pi^2}
\int_0^{\infty} \d q\, q^2 \left[ (e^{(E_i(q)-\mu_i)/T)}\pm 1)^{-1} - (e^{(E_i(q)+\mu_i)/T)}\pm 1)^{-1}\right] \nonumber\\
\equiv \frac{g_iT^{3}}{6}\times\begin{cases}
\beta\mu_i c_i(b) & \qquad\qquad{\rm bosons}\,,\\
\beta\mu_i c_i(f) & \qquad\qquad{\rm fermions}\,,
\end{cases}
\label{cdef}
\eeqn
where $n_{i}$ and $\bar{n}_{i}$ denote the equilibrium number density
of particle and antiparticle respectively,  $g_i$ counts the degrees
of freedom of the particle, $E_i(q)=\sqrt{q^2+m_i^2}$ where $m_i$ is the mass of particle $i$,
$\mu_i$ is the chemical potential of
the particle ($-\mu_i$ is the chemical potential of the antiparticle),
and $+1\,(-1)$ in the  denominator is for the case when the particle is a fermion (boson).
In the ultra relativistic limit $(T \gg m_i)$  the mass of the particle can be dropped.
In our analysis below we use the approximation of a weakly interacting
plasma  where $\beta\mu_i\ll1$, and  $\beta\equiv 1/T$ and one has
\begin{equation}
n_i-\bar{n}_i\  \sim \ \frac{g_iT^{3}}{6}\times\begin{cases}
2\beta\mu_i+\mathcal{O}\big((\beta\mu_i)^{3}\big) & \qquad\qquad{\rm bosons}\,,\\
\beta\mu_i+\mathcal{O}\big((\beta\mu_i)^{3}\big) & \qquad\qquad{\rm fermions}\,.
\end{cases}\label{NDandCP}
\end{equation}
In the limit where Eq.~(\ref{NDandCP}) holds we have $c_i(b)=2, c_i(f)=1$.
This limit is a useful approximation as it simplifies the
analysis of the chemical potentials that are needed in the generation of dark
matter. However, full analysis can be easily done by using the exact expression
of Eq.~\eqref{cdef}. We will discuss inclusion of these in Appendix~\ref{master}.
There the functions $c_i(b)$ and $c_i(f)$ introduced in Eq.~\eqref{cdef} will be found useful.
The mass of the dark matter is constrained by the experimental ratio of
dark matter to baryonic matter given in Eq.~(\ref{ratio}).
Defining $B$ to be the total baryon number in the Universe and $X$ to be
the total dark matter number, we obtain
\begin{equation}
\frac{\Omega_{\rm DM}}{\Omega_{\rm matter}}=\frac{X\cdot m_{\rm DM}}{B\cdot m_{\rm B}}\approx 5\,,
\end{equation}
so that the dark particle mass is given by
\begin{equation}
m_{\rm DM}\approx5\cdot\frac{B}{X}\cdot 1~{\rm GeV}\,.\label{DMmass0}
\end{equation}
Applying the general thermal equilibrium method~\cite{Harvey&Turner} (see also~\cite{Dreiner:1992vm}),
it is not difficult to express $B$ and $X$ in terms of the chemical potentials and
then find their ratio. We note one subtlety is that while $X$ and $B-L$ (where by $B-L$ we mean
the $B-L$ in the \sm sector) are conserved after the interaction in Eq.~(\ref{ASYINT}) decouples, $B$ is not.
Thus, for example, the top quark would drop out from the thermal bath at some temperature $T_t$ and
one must solve the new set of $\mu$ equations at $T<T_t$
which would affect the computation of $B$ although $B-L$ is conserved. Typically one takes
$T_t$ to be $M_t$ but it could be somewhat lower.
Specifically, as the temperature drops below $M_t\sim173$~GeV,
the top quark becomes semi-relativistic but
could still be involved in the thermal equilibrium constraints.
A precise determination of $T_t$ is out of the scope of this paper, and
here we simply assume that $T_t$ lies below $M_t$. Further, as the temperature
falls below the  temperature where sphaleron processes decouple, $B$ and $L$ would be separately
conserved down to the current temperatures. Thus the relevant $B$ to compute the dark matter mass
in Eq.~\eqref{DMmass0} would be the baryon number
below the sphaleron temperature which we label $B_{\rm final}$.  It is useful to express
$X$ and $B_{\rm final}$ in terms of $B-L$
so that $X=x(B-L)$ and $B_{\rm final}= b(B-L)$ where $b$
is to be determined later (see Eqs.~\eqref{B/B-L} and \eqref{B/B-L'}).
Thus, Eq.~\eqref{DMmass0} can be rewritten as
\begin{equation}
m_{\rm DM}\approx5\cdot\frac{b}{x}\cdot1~{\rm GeV}\,.\label{DMmass}
\end{equation}
\\

We will discuss six broad classes of models labeled Models A-F (see Table~\ref{tab_1}).
Models A,B,C are anchored in the \sm while
Model D is a two Higgs doublet (2HD) model.
For Models A and D, the asymmetry transfer interaction, of the form of Eq.~(\ref{ASYINT}),
is active only above the electroweak phase transition (EWPT) scale, i.e.,
$T_{{\rm int}}>T_{{\rm EWPT}}$ ($T_{{\rm EWPT}}\sim200-300~{\rm GeV}$
where the Higgs gets its VEV).
For Model B and C, the interaction which transfers the asymmetry
could be active \textit{also} below the EWPT scale, i.e., $T_{{\rm EWPT}}>T_{{\rm int}}$.
More specifically, in Model B we consider the temperature regime $T_{{\rm EWPT}}>T_{{\rm int}}>M_{t}$,
and in Model C we discuss $T_{t}>T_{{\rm int}}>M_{W}$ ($M_W$ is the mass of $W$ boson).
Similarly, we discuss models based on extensions of the MSSM.
Here we will focus on two cases; one of which is when $T_{\rm int}> M_{\rm SUSY}$ (Model E)
where $M_{\rm SUSY}$ is the (largest) soft breaking mass. In this case all the sparticles will be in the plasma.
The second case (Model F) corresponds to when the first two generations of sparticles (with mass $M_1$)
are heavy and drop out of the plasma (at some temperature $T_1<M_1$) while the third generation
sparticles, the gauginos, the Higgses and the Higgsinos (with mass $M_2\ll M_1$) remain in
the plasma. Thus for this case we have $T_1>T_{\rm int}>M_2>T_{{\rm EWPT}}$.
\\

These six cases are summarized in Table~\ref{tab_1}. There can be additional subcases
for these models corresponding to different choices of the $B-L$ transfer in Eq.~(\ref{ASYINT}).
\begin{table}[t!]
\begin{center}
\begin{tabular}{|c|c|c|}
\hline
Model A & \multirow{3} {*}{SM} & {$T_{{\rm int}}>T_{{\rm EWPT}}$} \tabularnewline
\cline{1-1} \cline{3-3}
Model B &  & \multicolumn{1}{c|}{$T_{{\rm EWPT}}>T_{{\rm int}}>M_{t}$}\tabularnewline
\cline{1-1} \cline{3-3}
Model C &  & \multicolumn{1}{c|}{$T_{t}>T_{{\rm int}}>M_{W}$}\tabularnewline
\hline
Model D & {2HD} & {$T_{{\rm int}}>T_{{\rm EWPT}}$} \tabularnewline
\hline
Model E & \multirow{2}{*}{MSSM} & \multicolumn{1}{c|}{$T_{\rm int}>M_{{\rm SUSY}}$}\tabularnewline
\cline{1-1} \cline{3-3}
Model F & & \multicolumn{1}{c|}{$T_1>T_{\rm int}>M_2>T_{{\rm EWPT}}$}\tabularnewline
\hline
\end{tabular}
\caption{\label{tab_1}A list of six models which allow for generation of asymmetric dark matter.
Models A,B,C are within the framework of the extensions of the standard model~(SM) while Model D is
an extension of a two Higgs doublet model~(2HD). Models E and F are
in the framework of an extension of the minimal supersymmetric standard model (MSSM).}
\end{center}
\end{table}

\section{Analysis in non-supersymmetric framework} \label{Sec:CPSM}
In this section we will determine  the
dark matter mass in terms of  the $B-L$ asymmetry in the
non-supersymmetric framework utilizing  Eq.~\eqref{DMmass}
for Models A-D. We will discuss
three different temperature regimes where the $B-L$ transfer takes place and then  deduce
a general formula for computing the asymmetric dark matter mass.
We note that the dark matter mass depends only on the $(B-L)$-charge of
the operator $\mathcal{O}_{\rm asy}$ that enters in
Eq.~\eqref{ASYINT} and not on other particulars of the interaction.
We will give several examples
of the operator $\mathcal{O}_{\rm asy}$ and compute the dark matter mass for them.

\subsection{$T>T_{{\rm EWPT}}$} \label{T>EWPT}
First we consider the case when the temperature is above the electroweak
phase transition  scale. In this case
the following fields are in  the relativistic
plasma in the early universe: three generations of left-handed lepton
doublets $L_i$ and quark doublets $q_{i}$, three generations
of right-handed charged leptons $e_{i}$ and up and down-type quarks
$u_{i}$ and $d_{i}$ ($i=1,2,3$), and number $\lambda_H$ of
complex Higgs doublets $H_i=(h_i^{+},h_i^{0})^T$.
Since the $Z$ boson and the photon couple to particle and anti-particle pairs
they have a vanishing chemical potential. Further,
in this  temperature regime, $SU(2)_{L}$ symmetry is unbroken, the $W$ and
$Z$ are part of the same gauge multiplet which requires that the
chemical potential of the $W$ vanishes.
The chemical potential of the gluon is zero and different color quarks carry the same chemical potential.
The flavor (CKM) mixing among quarks
ensures that the chemical potential of quarks in different generations
are equal. However for the lepton sector, there is no
such flavor mixing in the absence of neutrino masses~\cite{kt}.
Thus each of the lepton numbers ($L_e, L_{\mu}, L_{\tau}$) for the three
generations are separately conserved.
Our notation is as follows: $\mu_{L_i},\mu_{e_i}$
denote the chemical potentials of left-handed and right-handed leptons
while $\mu_{q_i},\mu_{u_i},\mu_{d_i}$ stand for the chemical potential of left-handed
and right-handed quarks. We assume that the chemical potential of all generations is
the same and thus drop the subscript $i$ and use $\mu_{H}$ for the chemical potential of
the Higgs doublets (we assume all the Higgs doublets have identical chemical potential).
\\

The Yukawa couplings
\begin{equation}
\mathcal{L}_{{\rm Yukawa}}=g_{e_i}\bar{L}_{i}He_{i}+g_{u_i}\bar{q}_{i}H^{c}u_{i}+g_{d_i}\bar{q}_{i}Hd_{i}
\label{2.1.1.a}
\end{equation}
yield the following relations among the
chemical potentials
\begin{gather}
\mu_{H}=\mu_{L}-\mu_{e}=\mu_{q}-\mu_{d}=\mu_{u}-\mu_{q}\,.
\label{2.1.1.b}
\end{gather}
Sphaleron processes ($\mathcal{O}_{{\rm sph}}\sim\prod_{i=1,2,3}q_{i}q_{i}q_{i}L_{i}$)
give us one additional relation,
\begin{equation}
3\mu_{q}+\mu_{L}=0\,.
\label{SPHEQ}
\end{equation}
The temperature at which sphaleron processes decouple
is estimated to be $T_{\rm Sph}$ $\sim$ $[80+54(m_h/120\GeV)]$ $\GeV$~\cite{AristizabalSierra:2010mv}.
It is very likely that $T_{\rm Sph}$ lies below
$T_{\rm EWPT}$, and thus the sphaleron processes are always active at $T>T_{\rm EWPT}$.
Finally, the hypercharge neutrality condition requires the total hypercharge of the Universe
 to be zero\footnote{\label{FN:YSM}
The hypercharge of the Universe used in deducing
Eq.~(\ref{2.1.1.d}) is computed as follows:
\begin{align}
Y=3\times\big[2\times3\times\tfrac{1}{3}\mu_{q}+3\times\tfrac{4}{3}\mu_{u}+3\times(-\tfrac{2}{3})\mu_{d}+2\times(-1)\mu_{L}+(-2)\mu_{e}\big]+2\times2\lambda_H\mu_{H}\,,\nonumber
\end{align}
where the  factor of 3 outside the first brace indicates summation  over
quark and lepton generations while inside the brace
the factor of 3 for quarks indicates summing over colors, the factor
of 2 for $q,L$ and $H$ counts  two fields inside the doublets,
and the additional factor of 2 for the Higgs is due to it being bosonic (see Eq.~\eqref{NDandCP}).}
\begin{align}
3\mu_{q}+6\mu_{u}-3\mu_{d}-3\mu_{L}-3\mu_{e}+2\lambda_H \mu_{H}=0\,.
\label{2.1.1.d}
\end{align}
Solving Eqs.~(\ref{2.1.1.a})-(\ref{2.1.1.d})
we can express all the chemical potentials
in terms of the chemical potential of one single field, e.g.,  $\mu_{L}$.
Specifically one finds for Model A with $\lambda_H=1$ (suppressing a factor of $\beta T^3/6$)
\begin{align}
B_{\rm A} & =3\times[2\mu_{q}+(\mu_{u}+\mu_{d})]=-4\mu_{L}\,,\\
L_{\rm A} & =3\times(2\mu_{L}+\mu_{e})=\frac{51}{7}\mu_{L}\,,
\end{align}
so that $(B-L)_{\rm A} = -\frac{79}{7} \mu_L$. And for Model D with $\lambda_H=2$ we have
\begin{align}
B_{\rm D}= -4\mu_{L}\,,\qquad L_{\rm D}  =\frac{15}{2}\mu_{L}\,,
\end{align}
and $(B-L)_{\rm D} = -\frac{23}{2} \mu_L$.

\subsection{$T<T_{{\rm EWPT}}$}
Now we consider the case when the temperature is below the EWPT scale.
After the Higgs gets its VEV, and the $SU(2)_{L}\times U(1)_Y$ symmetry is broken,   one has
$W^{\pm}$, $Z$, the photon and the Higgs scalar ($h$) as the physical particles in the
thermal bath. Again, since the $Z$ and the photon only couple to two particles
with opposite chemical potentials, their chemical potentials are zero.
For temperatures above the top quark mass,
the relativistic plasma
includes three generations of left-handed
and right-handed up-type and down-type quarks ($u_{iL}$, $u_{iR}$,
$d_{iL}$ and $d_{iR}$), three generations of left-handed leptons
($e_{iL}$ and $\nu_{i}$) and right-handed charged leptons ($e_{iR}$),
$i=1,2,3$.
As in Section~\ref{T>EWPT},
we will assume that the chemical potentials are generation
independent. Thus dropping the generation index we will
 use $\mu_{u_L},\mu_{u_R},\mu_{d_L},\mu_{d_R}$
to denote the chemical potentials of left-handed and right-handed
up-type and down-type quarks, $\mu_{e_L}$ and $\mu_{\nu}$ for left-handed
leptons, $\mu_{e_R}$ for right-handed charged leptons, $\mu_{W}$
for $W^{+}$, and $\mu_{h}$ for $h$.
\\

In the analysis below
we make the following approximations:
(1) At $T_{\rm EWPT}>T>M_t$, we still treat the top quark
as relativistic gas. (2) At $T_t>T>M_{W}$, we
treat the $W$ boson as relativistic
(all other particles, which have non-vanishing chemical potentials,
are very light so the limit in Eq.~\eqref{NDandCP} holds for them).
(3) We assume $T_t>T_{\rm Sph}$, i.e., the top quark drops out
of  the thermal bath before the sphaleron processes decouple.
\\

For $T<T_{{\rm EWPT}}$, the Yukawa couplings have the form
\begin{equation}
\mathcal{L}_{{\rm Yukawa}}=g_{e_i}\bar{e}_{iL}he_{iR}+g_{u_i}\bar{u}_{iL}hu_{iR}+g_{d_i}\bar{d}_{iL}hd_{iR}+h.c.\,,
\end{equation}
and since the Higgs boson is a real field and can couple to, for example,
both $\bar{e}_{iL}e_{iR}$ and $\bar e_{iR}{e}_{iL}$,
we get
\begin{gather}
0=\mu_{h}=\mu_{u_L}-\mu_{u_R}=\mu_{d_L}-\mu_{d_R}=\mu_{e_L}-\mu_{e_R}\,.
\label{h=0}
\end{gather}
Thus, the chemical potentials of left-handed and
right-handed quarks/charged leptons are equal.
The gauge interactions involving  $W$ bosons ($\mathcal{L}\sim W_{\mu}\bar{f}\gamma^{\mu}f$)
provide us the following relations,
\begin{align}
\mu_{W} & =\mu_{u_L}-\mu_{d_L}\qquad(W^{+}\leftrightarrow u_{L}+\bar{d}_{L})\,,\label{Y1} \\
\mu_{W} & =\mu_{\nu}-\mu_{e_L}\qquad(W^{+}\leftrightarrow \nu_{i}+\bar{e}_{iL})\,. \label{Y2}
\end{align}
The sphaleron processes give us one additional equation,
\begin{equation}
\mu_{u_L}+2\mu_{d_L}+\mu_{\nu}=0\,.\label{SPHEQ'}
\end{equation}
Since $SU(2)_{L}$ symmetry is broken below the EWPT scale, hypercharge
is no longer a good quantum number. Further, the neutrality of the Universe
now requires the total electrical charge to be zero\footnote{
The result of Eq.~(\ref{2.1.2.f}) follows from the computation of the total charge $Q$ which is given by
\begin{align}
Q=3\times\big[3\times\tfrac{2}{3}(\mu_{u_L}+\mu_{u_R})+3\times(-\tfrac{1}{3})(\mu_{d_L}+\mu_{d_R})+(-1)(\mu_{e_L}+\mu_{e_R})\big]+2\times3\mu_{W},
\end{align}
where again, the factors of 3 for fermions outside the big brace  indicates
summing over generations, the other factor of 3 for quarks stands
for summing over colors.  For the $W$ boson, 2 is the boson factor as given by
Eq.~(\ref{NDandCP}) and 3 is the degrees of freedom of $W$.
}
\begin{align}
2(\mu_{u_L}+\mu_{u_R}  + \mu_{W})-(\mu_{d_L}+\mu_{d_R} +\mu_{e_L}+\mu_{e_R})=0\,.
\label{2.1.2.f}
\end{align}
Solving the new set of equations one finds for Model B
\begin{align}
B_{\rm B} & =3\times[(\mu_{u_L}+\mu_{u_R})+(\mu_{d_L}+\mu_{d_R})]=-\frac{36}{7}\mu_{e}\,,\\
L_{\rm B} & =3\times(\mu_{e_L}+\mu_{e_R}+\mu_{\nu})=\frac{75}{7}\mu_{e}\,,
\end{align}
where we have expressed the results in terms of $\mu_e\equiv
\mu_{e_L}=\mu_{e_R}$, and $(B-L)_{\rm B} = -\frac{111}{7} \mu_e$.
\\

When the temperature drops below $T_t$, the top quark drops out
from the thermal bath, and we are left with just five flavors of quarks.
In this case ($T_t>T>M_{W}$) one must treat the first two
generations and the third generations separately.
For the first two generations the analysis of Eqs.~(\ref{h=0})-(\ref{SPHEQ'}) still holds.
For the remaining  third generation leptons, we assume  as before that the
chemical potentials  are identical to those for the first two generation leptons.
Further, we note that the charge current process
$W^{+}\leftrightarrow {u}_{L}+\bar{b}_{L}$ provides us with the relations
$\mu_{W}=\mu_{u_L}-\mu_{b_L}$ and $\mu_{b_L}=\mu_{d_L}$.
Thus we can treat Model C similar to Model B with only one modification to
the charge neutrality condition, which now becomes
\begin{align}
4(\mu_{u_L}+\mu_{u_R}) + 6\mu_{W}- 3(\mu_{d_L}+\mu_{d_R} +\mu_{e_L}+\mu_{e_R})=0\,.
\label{2.1.2.g}
\end{align}
Solving these equations we obtain for Model C
\begin{align}
B_{\rm C} & =2(\mu_{u_L}+\mu_{u_R})+3 (\mu_{d_L}+\mu_{d_R})=-\frac{90}{19}\mu_{e}\,, \label{BC} \\
L_{\rm C} & =3\times(\mu_{e_L}+\mu_{e_R}+\mu_{\nu})=\frac{201}{19}\mu_{e}\,, \label{LC}
\end{align}
and $(B-L)_{\rm C} = -\frac{291}{19} \mu_e$.
We note that the sphaleron processes will decouple below $T_{\rm Sph}$ as mentioned already.
Subsequently the baryon and lepton numbers would be separately conserved.
Eqs.~(\ref{h=0})-(\ref{Y2}), and (\ref{2.1.2.g})-(\ref{LC}) would remain valid at $T_{\rm Sph}> T > M_{W}$.
\\

Following our assumptions given earlier, the
top quark drops out of the thermal bath before sphaleron processes decouple.
After the sphaleron processes decouple, $B$ and $L$  would  be
separately conserved. In other words, the ratio of $B/(B-L)$ would
freeze as soon as the sphaleron processes are no longer active.
Thus, we obtain
\begin{equation}
b\,=\frac{B_{\rm final}}{B-L}=\left(\frac{B}{B-L}\right)_{\rm C}=\,\frac{30}{97}\,\approx\, 0.31\,.\label{B/B-L}
\end{equation}

\subsection{The AsyDM mass: non-SUSY case} \label{SMADMmass}
We discuss now in further detail the mechanism by which $B-L$ is transferred from the
standard model sector to the dark matter sector and the determination of the dark matter mass.
We consider the most general interaction which transfers the
$B-L$ asymmetry to dark matter at a high temperature:
\begin{equation}
\mathcal{L}_{{\rm asy}}^{{\rm SM}}=\frac{1}{M_{{\rm asy}}^{n}}X^{k}\mathcal{O}_{{\rm asy}}^{{\rm SM}}\,,
\label{3.3a}
\end{equation}
where the operator $\mathcal{O}_{{\rm asy}}^{{\rm SM}}$ is constructed from
the \sm fields,  has a $(B-L)$-charge $Q_{B-L}^{\mathcal{O}^{{\rm SM}}}$,
and $X$ is the dark particle and has a $(B-L)$-charge
$Q_{B-L}^{{\rm DM}}=-Q_{B-L}^{\mathcal{O}^{{\rm SM}}}/k$.\footnote{
The power of $X$ can only be 2 or greater to ensure the stability of the asymmetric dark matter.
}
\\

The parameterization of the asymmetric dark matter sector by the charge
$Q_{B-L}^{\rm DM}$  is useful and we will utilize it in our analysis below.
Also useful is the parameterization of the interactions in terms of the  number of
doublets and singlets that enter in $\mathcal{O}_{{\rm asy}}^{{\rm SM}}$,
i.e., $N_{q},N_{L},N_{H}$ numbers of $q,L,H$ doublets and $N_{u},N_{d},N_{e}$ numbers
of $u_{R},d_{R},e_{R}$  singlets which are all active above the EWPT scale.
Eq.~(\ref{3.3a}) leads to
the following constraints~\cite{imy}
\begin{alignat}{1}
N_{q}\mu_{q}+N_{L}\mu_{L}+N_{u}\mu_{u}+N_{d}\mu_{d}+N_{e}\mu_{e}+N_{H}\mu_{H}+k\mu_{X}&=0\,, \label{OPTcon1}\\
\tfrac{1}{3}N_{q}+\tfrac{1}{3}N_{u}+\tfrac{1}{3}N_{d}-N_{L}-N_{e}+kQ_{B-L}^{{\rm DM}}&=0\,, \label{OPTcon2}\\
\tfrac{1}{3}N_{q}+\tfrac{4}{3}N_{u}-\tfrac{2}{3}N_{d}-N_{L}-2N_{e}+N_{H}&=0\,. \label{OPTcon3}
\end{alignat}
Here Eq.~(\ref{OPTcon1}) arises from the $\mu$ equilibrium of Eq.~(\ref{3.3a}),
Eq.~(\ref{OPTcon2}) arises from the total $(B-L)$-charge conservation of the interaction,
and Eq.~(\ref{OPTcon3}) arises from the hypercharge conservation and the condition that the
asymmetric dark matter must have zero hypercharge.
Together with Eqs.~\eqref{2.1.1.b}-\eqref{2.1.1.d}, for Model A we obtain
\begin{equation}
\mu_{X}^{{\rm A}}=-\frac{11}{7}Q_{B-L}^{{\rm DM}}\mu_{L}\,.
\end{equation}
If $X$ is fermionic dark matter~(FDM), we find,
\begin{equation}
x_{{\rm A}}=\frac{X_{{\rm A}}}{(B-L)_{{\rm A}}}=\frac{k\mu_{X}^{{\rm A}}}{-\tfrac{79}{7}\mu_{L}}=-\frac{11}{79}Q_{B-L}^{\mathcal{O}^{{\rm SM}}}\,.
\end{equation}
Using Eqs.~\eqref{DMmass} and \eqref{B/B-L}, we obtain
\begin{equation}
m_{{\rm FDM}}^{{\rm A}}\approx-\frac{11.11~{\rm GeV}}{Q_{B-L}^{\mathcal{O}^{{\rm SM}}}}\,.\label{DMMFA}
\end{equation}
Similarly, for Model D with two Higgs doublets, we have
\begin{equation}
\mu_{X}^{{\rm D}}=-\frac{3}{2}Q_{B-L}^{{\rm DM}}\mu_{L}\,,
\end{equation}
so that
\begin{equation}
m_{{\rm FDM}}^{{\rm D}}\approx-\frac{11.86~{\rm GeV}}{Q_{B-L}^{\mathcal{O}^{{\rm 2HD}}}}\,.\label{DMMFD}
\end{equation}
\\

If the $B-L$ transfer interaction is also active below the EWPT scale, the treatment is similar.
Assuming
$\mathcal{O}_{{\rm asy}}^{{\rm SM}}$ has $N_{u},N_{d},N_{e},N_{\nu},N_{W}$
numbers of $u,d,e,\nu,W^{+}$ fields
 and recalling  that at
$T<T_{{\rm EWPT}}$, the left-handed and right-handed quarks and charged
leptons have the same chemical potentials, one finds the following constraints
\begin{alignat}{1}
N_{u}\mu_{u}+N_{d}\mu_{d}+N_{e}\mu_{e}+N_{\nu}\mu_{\nu}+N_{W}\mu_{W}+k\mu_{X} & =0\,,\\
\tfrac{1}{3}N_{u}+\tfrac{1}{3}N_{d}-N_{e}-N_{\nu}+k Q_{B-L}^{{\rm DM}} & =0\,,\\
\tfrac{2}{3}N_{u}-\tfrac{1}{3}N_{d}-N_{e}+N_{W} & =0\,.
\end{alignat}
We note that the last condition is from the charge neutrality of  the operator $\mathcal{O}_{{\rm asy}}^{{\rm SM}}$.
Together with Eqs.~\eqref{h=0}-\eqref{2.1.2.f}, we obtain for Model B,
\begin{equation}
\mu_{X}^{{\rm B}}=-\frac{11}{7}Q_{B-L}^{{\rm DM}}\mu_{e}\,.
\end{equation}
The fermionic dark matter mass in this model reads
\begin{equation}
m_{{\rm FDM}}^{{\rm B}}\approx-\frac{15.60~{\rm GeV}}{Q_{B-L}^{\mathcal{O}^{{\rm SM}}}}\,.\label{DMMFB}
\end{equation}
For Model C where the top quark is  out of the thermal bath, we find
\begin{equation}
\mu_{X}^{{\rm C}}=-\frac{29}{19}Q_{B-L}^{{\rm DM}}\mu_{e}\,.
\end{equation}
and
\begin{equation}
m_{{\rm FDM}}^{{\rm C}}\approx-\frac{15.52~{\rm GeV}}{Q_{B-L}^{\mathcal{O}^{{\rm SM}}}}\,.\label{DMMFC}
\end{equation}
\\

Now we consider the simplest example of the $B-L$ transfer interaction ($Q_{B-L}^{\mathcal{O}^{{\rm SM}}}=-1$)
\begin{equation}
\mathcal{L}_{{\rm asy}}=\frac{1}{M_{{\rm  asy}}^{3}}\psi^{3}LH\,,\label{ASYInt1}
\end{equation}
where $\psi$ is the fermionic dark matter (which carries a lepton number of $-1/3$)
and $\psi^{3}\equiv\bar{\psi}^{c}\psi\bar{\psi}^{c}$.
If this interaction is only active above the EWPT scale then
the dark matter masses
 in Models A and D,
and more appropriately in Models $\rm A_1$ and $\rm D_1$ since the interaction of Eq.~\eqref{ASYInt1}
is being used (see Table~\ref{tab_2} which also includes a list of additional interactions), are computed to be
\begin{equation}
m_{\psi} = 11.11\gev\qquad{\rm Model~A_1}\,;\qquad m_{\psi} = 11.86\gev\qquad{\rm Model~D_1}\,.
\end{equation}
If this interaction is also active below the EWPT scale, the dark matter masses
in Models B and C are:
\begin{equation}
m_{\psi} = 15.60\gev\qquad{\rm Model~B_1}\,;\qquad m_{\psi} = 15.52\gev\qquad{\rm Model~C_1}\,.
\end{equation}
Further, applying Eq.~(\ref{MST}) and the
bounds in Table~\ref{tab_1}
one can estimate the mass scales for these interactions:
\begin{gather}
M_{{\rm asy}}^{\rm A_1/D_1}\gtrsim1.2\times10^{5}~{\rm GeV}\,, \label{sm_mA1} \\
1.2\times10^{5}~{\rm GeV}\gtrsim M_{{\rm asy}}^{\rm B_1} \gtrsim 0.9\times10^{5}~{\rm GeV}\,, \label{sm_mB1} \\
0.9\times10^{5}~{\rm GeV} > M_{{\rm asy}}^{\rm C_1}\gtrsim 0.4\times10^{5}~{\rm GeV}\,. \label{sm_mC1}
\end{gather}
\\

In the analysis above we focused on  asymmetric fermionic dark matter.
For bosonic dark matter, the masses would be half the fermionic ones, c.f., Eq.~\eqref{NDandCP}.
As an example, we consider now an interaction with a higher dimensional operator $\mathcal{O}^{{\rm SM}}_{\rm asy}$:
\begin{equation}
\mathcal{L}_{{\rm asy}}=\frac{1}{M_{{\rm asy}}^{n}}X^{2}(LH)^{2}.\label{LH^2}
\end{equation}
In this case, the dark matter  could be either a fermion ($X=\psi$, $n=4$)
or a boson ($X=\phi$, $n=3$).
This interaction gives rise to Models ${\rm A_2}$-${\rm D_2}$
and Models ${\rm A_3}$-${\rm D_3}$.
As examples, for Models ${\rm A_2}$ and ${\rm A_3}$ where $T_{{\rm int}}>T_{{\rm {\rm EWPT}}}$,
applying Eq.~\eqref{DMMFA} we find that
the dark matter masses  are
\begin{equation}
m_{\psi}=5.55~{\rm GeV}\qquad{\rm Model~A_2}\,;\qquad m_{\phi}=2.78~{\rm GeV}\qquad{\rm Model~A_3}\,.
\end{equation}
\\

We explain now briefly the equality of asymmetric dark mass for the Models $\rm A_1, A_4, A_5, A_6$.
From Eq.~(\ref{ASYINT}) we can write
\beqn
\mu_{\mathcal{O}_{\rm DM}}+ \mu_{\mathcal{O}_{\rm asy}^{\rm SM}}=0\,.
\eeqn
For Models $\rm A_1$,$\rm A_4$-$\rm A_6$ we have
\begin{align}
LH \; ({\rm A_1}): & \qquad \mu_{\mathcal{O}_{\rm asy,1}^{\rm SM}}= \mu_L + \mu_H\,, \\
LLe^c \; ({\rm A_4}): & \qquad \mu_{\mathcal{O}_{\rm asy,4}^{\rm SM}}= 2\mu_L - \mu_{e}\,, \\
Lqd^c \; ({\rm A_5}): & \qquad \mu_{\mathcal{O}_{\rm asy,5}^{\rm SM}}= \mu_L+ \mu_q - \mu_{d}\,, \\
u^cd^cd^c \; ({\rm A_6}): & \qquad \mu_{\mathcal{O}_{\rm asy,6}^{\rm SM}}= - \mu_{u} - 2 \mu_{d}\,.
\end{align}
From the $\mu$ equations Eqs.~\eqref{2.1.1.b} and \eqref{SPHEQ}, it is easy to see that
\beqn
\mu_{\mathcal{O}_{\rm asy,1}^{\rm SM}} = \mu_{\mathcal{O}_{\rm asy,4}^{\rm SM}}
= \mu_{\mathcal{O}_{\rm asy,5}^{\rm SM}} = \mu_{\mathcal{O}_{\rm asy,6}^{\rm SM}}\,.
\label{3.z}
\eeqn
Eq.~(\ref{3.z}) implies that the dark matter has the same mass
for the Models $\rm A_1$,$\rm A_4$-$\rm A_6$. Similar analysis holds for Models
$\rm B_1$,$\rm B_4$-$\rm B_6$, $\rm C_1$,$\rm C_4$-$\rm C_6$ and $\rm D_1$,$\rm D_4$-$\rm D_6$.
\\

\begin{table}
\begin{center}
\begin{tabular}{|c|c|c|c|c|c|c|c|c|}
\hline
$\frac{1}{M^{n}}X^k\mathcal{O}_{{\rm asy}}^{{\rm SM}}$
& Model  & DM Mass & Model & DM Mass & Model & DM Mass & Model & DM Mass \tabularnewline
\hline
\hline
$\frac{1}{M^{3}}\psi^{3}LH$  &  ${\rm A_1}$ & 11.11~GeV &${\rm B_1}$ & 15.60~GeV & ${\rm C_1}$ & 15.52~GeV & ${\rm D_1}$ & 11.86~GeV \tabularnewline
$\frac{1}{M^{4}}\psi^{2}(LH)^{2}$ & ${\rm A_2}$ & 5.55~GeV &${\rm B_2}$ & 7.80~GeV & ${\rm C_2}$ & 7.76~GeV & ${\rm D_2}$ & 5.93~GeV \tabularnewline
$\frac{1}{M^{3}}\phi^{2}(LH)^{2}$ & ${\rm A_3}$ & 2.78~GeV &${\rm B_3}$ & 3.90~GeV & ${\rm C_3}$ & 3.88~GeV & ${\rm D_3}$ & 2.96~GeV \tabularnewline
$\frac{1}{M^{5}}\psi^{3}L L e^c$ & ${\rm A_4}$ & 11.11~GeV &${\rm B_4}$ & 15.60~GeV & ${\rm C_4}$ & 15.52~GeV & ${\rm D_4}$ & 11.86~GeV \tabularnewline
$\frac{1}{M^{5}}\psi^{3}L q d^c$ & ${\rm A_5}$ & 11.11~GeV &${\rm B_5}$ & 15.60~GeV & ${\rm C_5}$ & 15.52~GeV & ${\rm D_5}$ & 11.86~GeV \tabularnewline
$\frac{1}{M^{5}}\psi^{3} u^{c} d^{c} d^{c}$ & ${\rm A_6}$ & 11.11~GeV &${\rm B_6}$ & 15.60~GeV & ${\rm C_6}$ & 15.52~GeV & ${\rm D_6}$ & 11.86~GeV \tabularnewline
\hline
\end{tabular}
\caption{\label{tab_2}A display of the various interactions
that allow a transfer of the $B-L$ asymmetry from
the standard model sector to the dark matter sector.}
\end{center}
\end{table}

We summarize all our results in Table~\ref{tab_2}, where we list the dark
matter mass for the various interactions\footnote{
In the first column of Table~\ref{tab_2}, $L,H$ and $q$ stand for $SU(2)_L$ doublets as
discussed in $T>T_{\rm EWPT}$ regime (Model A and D).
When the temperature drops below EWPT scale (Model B and C), since $SU(2)_L$ symmetry is broken,
these interactions should be rewritten in terms of the contents of the original doublets.
We omit this step for simplicity.
} that can transfer the $B-L$ asymmetry from the
standard model sector to the dark matter sector.
We  note that for the first five interactions,
the dark matter  carries lepton  number,  while for
the last one, it carries a baryon number.

\subsection{The AsyDM mass: including the right-handed neutrinos} \label{SMADMmassRHN}

In the analysis above we used the framework of the standard model where we
have no right-handed neutrinos and the neutrinos are assumed massless.  The nature
of neutrino masses is currently not known, i.e., whether they are Majorana or Dirac,
but in the context of  a gauged $L_{\mu}-L_{\tau}$ symmetry it is more
natural for the neutrinos to have Dirac masses which implies that we  introduce right-handed neutrinos, one for
each generation.  We discuss now the effect of this inclusion on the analysis, i.e., on the $\mu$
equations, on the  $B/(B-L)$ ratio and thus on the DM mass.
\\

Since the right-handed neutrino $\nu_{R}$ has 0 hypercharge and 0 electrical charge, it does not affect
the neutrality conditions (such as Eq.~\eqref{2.1.1.d} or Eq.~\eqref{2.1.2.f}).
$\nu_{R}$ is only involved in one interaction
$\mathcal{L}\sim\bar{L}_i H^c \nu_{iR}$ before the electroweak phase transition
(or $\mathcal{L}\sim\bar{\nu}_{iL} h \nu_{iR}$ after EWPT),
which gives us $\mu_H=\mu_{\nu_{iR}}-\mu_{L_i}$ before EWPT (or $\mu_{\nu_{iR}}=\mu_{\nu_{iL}}$ after EWPT).
Thus the only change would be the total
lepton number since $\nu_{R}$ carries lepton number 1.
By including the right-handed neutrinos, a reanalysis gives the following formulas
\begin{align}
m_{{\rm FDM}}^{{\rm A'}}\approx-\frac{12.12~{\rm GeV}}{Q_{B-L}^{\mathcal{O}^{{\rm SM}}}}\,,\qquad
m_{{\rm FDM}}^{{\rm D'}}\approx-\frac{12.70~{\rm GeV}}{Q_{B-L}^{\mathcal{O}^{{\rm 2HD}}}}\,, \label{DMMFAD'}\\
m_{{\rm FDM}}^{{\rm B'}}\approx-\frac{15.58~{\rm GeV}}{Q_{B-L}^{\mathcal{O}^{{\rm SM}}}}\,,\qquad
m_{{\rm FDM}}^{{\rm C'}}\approx-\frac{15.52~{\rm GeV}}{Q_{B-L}^{\mathcal{O}^{{\rm SM}}}}\,, \label{DMMFBC'}
\end{align}
where we use a prime to denote all the models with the right-handed neutrinos.
With the inclusion of right-handed neutrinos in the thermal bath
$b'$ is determined to be  (c.f. Eq.~\eqref{DMmass})
\begin{equation}
b'\,=\frac{B_{\rm final}}{B-L}=\left(\frac{B}{B-L}\right)_{\rm C'}=\,\frac{5}{21}\,\approx\, 0.24\,.\label{B/B-L'}
\end{equation}
With above formulas, we compute the dark matter masses for the same $B-L$ transfer interactions
displayed in Table~\ref{tab_2}, and they are collected in Table~\ref{tab_3}.
We note that inclusion of right-handed neutrinos generates less than a 10\% effect at most and
no effect for Model C.\footnote{
The reason for this is simple, i.e.,  using Eq.~\eqref{DMmass0} and recalling  the fact
that $B_{\rm final}=B_{\rm C}=B_{\rm C'}$, since the inclusion of
right-handed neutrinos does not change the total baryon number, we have
\begin{equation}
m_{{\rm DM}}^{{\rm C'}}=5\cdot\frac{B_{\rm final}}{X}\cdot 1~{\rm GeV}
=5\cdot\frac{B_{\rm C'}}{X}\cdot 1~{\rm GeV}=5\cdot\frac{B_{\rm C}}{X}\cdot 1~{\rm GeV}=m_{{\rm DM}}^{{\rm C}}\,.\nonumber
\end{equation}
}

\begin{table}
\begin{center}
\begin{tabular}{|c|c|c|c|c|c|c|c|c|}
\hline
$\frac{1}{M^{n}}X^k\mathcal{O}_{{\rm asy}}^{{\rm SM}}$
& Model  & DM Mass & Model & DM Mass & Model & DM Mass & Model & DM Mass \tabularnewline
\hline
\hline
$\frac{1}{M^{3}}\psi^{3}LH$  &  ${\rm A_1'}$ & 12.12~GeV &${\rm B_1'}$ & 15.58~GeV & ${\rm C_1'}$ & 15.52~GeV & ${\rm D_1'}$ & 12.70~GeV \tabularnewline
$\frac{1}{M^{4}}\psi^{2}(LH)^{2}$ & ${\rm A_2'}$ & 6.06~GeV &${\rm B_2'}$ & 7.79~GeV & ${\rm C_2'}$ & 7.76~GeV & ${\rm D_2'}$ & 6.35~GeV \tabularnewline
$\frac{1}{M^{3}}\phi^{2}(LH)^{2}$ & ${\rm A_3'}$ & 3.03~GeV &${\rm B_3'}$ & 3.90~GeV & ${\rm C_3'}$ & 3.88~GeV & ${\rm D_3'}$ & 3.18~GeV \tabularnewline
$\frac{1}{M^{5}}\psi^{3}L L e^c$ & ${\rm A_4'}$ & 12.12~GeV &${\rm B_4'}$ & 15.58~GeV & ${\rm C_4'}$ & 15.52~GeV & ${\rm D_4'}$ & 12.70~GeV \tabularnewline
$\frac{1}{M^{5}}\psi^{3}L q d^c$ & ${\rm A_5'}$ & 12.12~GeV &${\rm B_5'}$ & 15.58~GeV & ${\rm C_5'}$ & 15.52~GeV & ${\rm D_5'}$ & 12.70~GeV \tabularnewline
$\frac{1}{M^{5}}\psi^{3} u^{c} d^{c} d^{c}$ & ${\rm A_6'}$ & 12.12~GeV &${\rm B_6'}$ & 15.58~GeV & ${\rm C_6'}$ & 15.52~GeV & ${\rm D_6'}$ & 12.70~GeV \tabularnewline
\hline
\end{tabular}
\caption{\label{tab_3}A display of the various interactions
that allow a transfer of the $B-L$ asymmetry from
the standard model sector (including the right-handed Dirac neutrinos) to the dark matter sector.}
\end{center}
\end{table}

\section{Analysis in supersymmetric framework} \label{Sec:CPMSSM}
We now consider the analysis  in a supersymmetric framework specifically within an
extended MSSM. Since the supersymmetric case can have its own dark matter candidate, i.e.,
the neutralino, the relic abundance of the neutralino must be depleted. For this reason, we only consider the parameter space where  relic density of the neutralino is much smaller than the WMAP value
for cold dark matter~(CDM) and is thus only a subdominant component. Below we discuss two regimes, one where
$T_{\rm int}> M_{\rm SUSY}$ and the other where $T_1>T_{\rm int}>M_2>T_{{\rm EWPT}}$.

\subsection{$T>M_{{\rm SUSY}}$}
In this regime since the temperature is above the SUSY breaking scale
all sparticle masses must be included in the $\mu$ equations.
This case is very
similar to the discussion of $T>T_{{\rm EWPT}}$ in the \sm framework,
except this time our particle spectrum includes all the \sm particles, the extra Higgses
as well as the sparticles. For brevity we will use the
same symbols for the chemical potentials, though now they stand for  not
only the \sm fields, but also their super-partners. The chemical potential
equations obtained from Yukawa couplings and sphaleron processes remain
the same. The only  equation modified would be the hypercharge equation,
which becomes\footnote{
The hypercharge of the Universe for the case when $T> T_{\rm SUSY}$ is given by
\begin{align}
Y=3\times\big\{3\times\big[2\times3\times\tfrac{1}{3}\mu_{q}+3\times\tfrac{4}{3}\mu_{u}+3\times(-\tfrac{2}{3})\mu_{d}+2\times(-1)\mu_{L}+(-2)\mu_{e}\big]+2\times(\mu_{H_{u}}-\mu_{H_{d}})\big\},\nonumber
\end{align}
where the counting is similar to discussion in footnote \ref{FN:YSM}. The Higgs mixing term in the
superpotential, i.e.,
$W=\mu H_{u}H_{d}$ indicates $\mu_{H_{u}}+\mu_{H_{d}}=0$,  and
so we define $\mu_{H}\equiv\mu_{H_{u}}=-\mu_{H_{d}}$.}
\begin{align}
3\mu_{q}+6\mu_{u}-3\mu_{d}-3\mu_{L}-3\mu_{e}+2\mu_{H}=0\,.
\label{2.2.1.a}
\end{align}
Solving the  chemical potential equations, we find that for Model E the total baryon and lepton numbers are
given by
\begin{align}
B_{\rm E} & =3\times3\times[2\mu_{q}+(\mu_{u}+\mu_{d})]=-12\mu_{L}\,,\\
L_{\rm E} & =3\times3\times(2\mu_{L}+\mu_{e})=\frac{153}{7}\mu_{L}\,,
\end{align}
so that $(B-L)_{\rm E}=-\frac{237}{7}\mu_L$. Note that in the above equations,
the extra factor of $3 = 1 + 2$ (compare to the \sm case) takes into account
the contributions of both fermions and bosons from the superfields, c.f. Eq.~\eqref{NDandCP}.

\subsection{$T_1>T>M_2>T_{{\rm EWPT}}$}
Here we consider two soft breaking mass scales $M_1$ and $M_2$
where $M_1 \gg M_2$. When temperature drops below $T_1$, all
the super-particles with masses greater than $M_1$ would drop out of the thermal bath.
We assume that this is the case for the first two generations
of squarks and sleptons. Similar to Model C, we simply assume here that
these super-particles would drop out of the thermal bath at $M_1>T_1>M_2$.
Thus the only super-particles remaining in the
thermal bath are the third generation sparticles,
the gauginos, the Higgses and the Higgsinos.
We make the approximation that these particles are relativistic at $T_1>T>M_2$.
This case is labeled Model F.
Following the analysis of Eq.~(\ref{2.2.1.a}) we find that the vanishing
of the hypercharge for Model F gives
\begin{align}
5\mu_{q}+10\mu_{u}-5\mu_{d}-5\mu_{L}-5\mu_{e}+6\mu_{H}&=0\,.
\end{align}
Solving the $\mu$-equations, we obtain
\begin{align}
B_{\rm F} & =(3\times1+2)\times[2\mu_{q}+(\mu_{u}+\mu_{d})]=-\frac{20}{3}\mu_{L}\,,\\
L_{\rm F} & =(3\times1+2)\times(2\mu_{L}+\mu_{e})=\frac{485}{39}\mu_{L}\,,
\end{align}
and $(B-L)_{\rm F}=-\frac{745}{39}\mu_L$.

\subsection{The AsyDM mass: SUSY case}
The supersymmetric interactions which transfer $B-L$ asymmetry typically have a different form than the
ones in the non-supersymmetric case. The most general interaction that
transfers $B-L$ to the dark sector for the MSSM case is
\begin{equation}
W_{{\rm asy}}=\frac{1}{M^n_{{\rm asy}}}X^k \mathcal{O}_{\rm asy}^{\rm MSSM}\,,\label{asysusy}
\end{equation}
where the dark matter superfield $X= (\phi_X, \psi_X)$ with $\phi_X$
as the bosonic and $\psi_X$ as the fermionic component.
Now the following possibilities arise in terms of dark matter.
First, after soft breaking if $\phi_X$ and $\psi_X$ have a similar mass,  both of them are stable,
and could be  dark matter candidates. Next, consider the case where one of the components has a
much larger mass than the other and would decay into the lighter one. In this case we have
two possibilities: either $\phi_X$ is heavier than $\psi_X$ so that
$\phi_X \rightarrow \psi_X + \tilde{\chi}^{\rm St}$ (where $\tilde {\chi}^{\rm St}$ is
a \st neutralino)
in which case $\psi_X$ is the dark matter candidate,
or $\psi_X$ is heavier than $\phi_X$ so that
$\psi_X \rightarrow \phi_X + \tilde{\chi}^{\rm St}$ in which case $\phi_X$ is the dark matter candidate
(The possibility that either $\tilde {\chi}^{\rm St}$ or the MSSM neutralino is a dark matter
candidate is discussed in Section~\ref{Sec:ASY_MSSM}).
For either of these three cases, when computing the total dark particle number from Eq.~\eqref{NDandCP},
we need to multiply by an additional factor of 3, since both bosonic and fermionic components of
the dark matter superfield would contribute.
But for concreteness in our analysis we will assume that $\psi_X$ is lighter than $\phi_X$
and thus would be the asymmetric dark matter.
\\

Applying the same method we used in Section~\ref{SMADMmass}, we find
\begin{align}
m_{{\rm DM}}^{{\rm E}}\approx-\frac{11.11~{\rm GeV}}{Q_{B-L}^{\mathcal{O}^{{\rm MSSM}}}}\,,\qquad
m_{{\rm DM}}^{{\rm F}}\approx-\frac{6.51~{\rm GeV}}{Q_{B-L}^{\mathcal{O}^{{\rm MSSM}}}}\,.\label{DMMFEF}
\end{align}
Thus for the $B-L$ transfer interactions with $Q_{B-L}^{\mathcal{O}^{{\rm MSSM}}}=-1$,
where $\mathcal{O}_{\rm asy}^{\rm MSSM}$ can be $LH_u$, $LLe^c$, $Lqd^c$, or $u^c d^c d^c$,
the dark particle masses are
\begin{align}
m_{X} = 11.11~{\rm GeV}\qquad{\rm Model~E}\,;\qquad m_{X} = 6.51~{\rm GeV}\qquad{\rm Model~F}\,.
\end{align}
For the case $W_{\rm asy}=\frac{1}{M_{\rm asy}^3} X^2 (LH_u)^2$ with $Q_{B-L}^{\mathcal{O}^{{\rm MSSM}}}=-2$,
which we will discuss  in Section~\ref{Sec:ASY_MSSM},
the dark particle masses are
\begin{align}
m_{X} = 5.55~{\rm GeV}\qquad{\rm Model~E}\,;\qquad m_{X} = 3.25~{\rm GeV}\qquad{\rm Model~F}\,,
\end{align}
and using  Eq.~(\ref{MST}) one finds
\begin{equation}
M_{{\rm asy}}^{\rm E} \gtrsim  3.7\times 10^{5}~{\rm GeV}\,.
\end{equation}
\\

If we include the right-handed  neutrinos in the supersymmetric framework,
the analysis is similar, and Eq.~\eqref{DMMFEF} becomes
\begin{align}
m_{{\rm DM}}^{{\rm E'}}\approx-\frac{12.12~{\rm GeV}}{Q_{B-L}^{\mathcal{O}^{{\rm MSSM}}}}\,,\qquad
m_{{\rm DM}}^{{\rm F'}}\approx-\frac{6.99~{\rm GeV}}{Q_{B-L}^{\mathcal{O}^{{\rm MSSM}}}}\,.\label{DMMFEF'}
\end{align}

\section{Asymmetric dark matter in a \st extension of the SM} \label{Sec:ASY_SM}
As discussed in the Introduction, one of the major problems for an acceptable AsyDM model is
to have an efficient mechanism for the annihilation of dark matter that is produced thermally.
In general one has
\beqn
\Omega_{\rm DM} = \Omega_{\rm DM}^{\rm asy} + \Omega_{\rm DM}^{\rm sym}\,,
\label{6.1}
\eeqn
where $\Omega_{\rm DM}^{\rm asy}$ is the relic density of asymmetric dark matter (which carries a
nonzero $(B-L)$-charge) and $\Omega_{\rm DM}^{\rm sym}$ is the relic density of dark matter which is produced thermally.
 For the asymmetric dark matter to be the dominant component, one must significantly
deplete the symmetric component of dark matter. Specifically we will use the criteria that
$\Omega_{\rm DM}^{\rm sym}\,/\,\Omega_{\rm DM} <0.1$.\footnote{
The analysis of previous sections was based on the assumption $\Omega_{\rm DM}^{\rm asy}\,/\,\Omega_{\rm matter}\approx 5$. Inclusion of a small contribution (i.e.,  $\leq 10\%$) of symmetric component to dark matter will
proportionately affect the determination of the dark matter mass. It is straightforward to take account of
this contribution but we do not carry it out  explicitly as it is a relatively small effect.}
Thus we investigate if the symmetric component of dark matter
produced by thermal processes can be annihilated efficiently.
We accomplish this via the exchange of a gauge field using
the \st formalism where the gauge field couples to $L_{\mu}- L_{\tau}$.
\\

For illustration let us consider Model $\rm A_1$, which
is governed by the interaction Eq.~\eqref{ASYInt1} operating at $T_{\rm int}>T_{\rm EWPT}$.
The corresponding dark matter mass is
11.11~GeV. Further, we require the dark matter
particles $\psi$ to have a non-vanishing $\mu$ or $\tau$ lepton number.
The total Lagrangian is given by
\beqn
\Lag =
\Lag_{\rm SM} + \Lag_{ U(1)}+  \Lag_{\rm St}~,
\eeqn
%%%%
where $\Lag_{ U(1)}$ is the kinetic energy for
the gauge field for the $L_{\mu}-L_{\tau}$ symmetry,
and for $\Lag_{\rm St}$ we assume the following form:
\beqn
\Lag_{\rm St}= - \frac{1}{2} (M_C C_{\mu}+  \partial_{\mu} \sigma)^2~.
\label{mass3}
\eeqn
In the unitary gauge the massive vector boson field will be called $\zp$ and its
  interaction  with fermions in the theory is given by
\beqn
\Lag_{\rm int} = \coup[\psi] \bar{\psi}  \gamma^{\mu} \psi C_{\mu}  + \coup[f] \bar f \gamma^{\mu} f {C_{\mu}}~,
\label{smint}
\eeqn
%%%%%%%%%%%%%%%%
where $f$ runs over $\mu$ and $\tau$ families and  $Q_C^{\mu}=-Q^{\tau}_C$.

\subsection{Resonant annihilation of symmetric dark matter}
We discuss now the details of
the annihilation of the symmetric component of
dark  matter. We will show that the relic density for such dark matter
can be reduced significantly below the WMAP value with resonant annihilation via the $Z'$ pole,
i.e., via the process
$\psi\bar{\psi}\to\zp\to f\bar{f}$.\footnote{
While the thermal dark matter can annihilate into second and third generation leptons
at the tree-level,  such an annihilation into the first generation leptons can come about only
at the loop level involving the second and third generation leptonic loops.
Thus the annihilation of thermal dark matter into first generation leptons
is significantly suppressed relative to the annihilation into the second and the third generation
leptons.}
Thus, by using  Eq.~\eqref{smint} one can compute the $\psi \bar{\psi} \to f\bar f$
annihilation cross section and using the Breit-Wigner form for a resonance one has
\begin{eqnarray}
\sigma_{\psi\bar \psi\to f\bar f} &=& a_{\psi}
\left|\left({s-M_{\zp}^2+i\Gamma_{\zp} M_{\zp}}\right)\right|^{-2}\,,\\
a_{\psi} &=& \frac{\beta_f (\tfrac{1}{2}g^2_CQ_C^{\psi} Q^f_C)^2} {64 \pi s \beta_{\psi}}
\left[ s^2 (1+ \frac{{1}}{3} \beta_f^2\beta_{\psi}^2)
+ 4M_{\psi}^2 (s-2m_f^2) + {4} m_f^2 (s+ 2 M_{\psi}^2)\right]\,,
\end{eqnarray}
where $\beta_{f,\psi} = (1-4m_{f,\psi}^2/s)^{1/2}$.
The relevant partial $\zp$ decay {widths are} given by
\beqn
\Gamma(\zp\to f \bar f ) &= &
\left(\coup[f]\right)^2 r_f \frac{M_{\zp}}{12\pi},\quad f=\mu, \nu_{\mu}, \tau,\nu_{\tau} \label{aa}\\
\Gamma(\zp\to \psi \bar \psi )&=&  \left(\coup[\psi]\right)^2\frac{M_{\zp}}{12\pi}
\left(1+\frac{2M_{\psi}^2}{M^2_{\zp}}\right)\left(1-\frac{4M_{\psi}^2}{M^2_{\zp}}\right)^{1/2}
\Theta\left(M_{\zp} - 2 M_{\psi}\right)~,
\label{aaa}
\eeqn
where $r_f=1$ for $f=\mu,\tau$ and $r_f=1/2$ for $f=\nu_{\mu}, \nu_{\tau}$.
A constraint on $g_C$ comes from the contribution of the $\zp$ to
$g_{\mu}-2$~\cite{pdgrev,gmuon2},
which is given by
\beqn
\Delta (g_{\mu}-2) = \left(\coup[\mu]\right)^2 \frac{m_{\mu}^2}{ 6 \pi^2 M_{\zp}^2}~.
\eeqn
In the analysis here we impose the constraint
that the $\zp$ boson  contribution be less than  the
experimental ($4\sigma$) deviation  of
$\Delta a_{\mu}\equiv\Delta \big((g_{\mu}-2)/2\big)= \left(3.0\pm0.8\right)\times 10^{-9}$~\cite{pdgrev,gmuon2},
which is the constraint commonly adopted in analysis of supergravity based models.
\\

The relic densities of $\psi$ and $\bar \psi$ are governed by the Boltzmann equations and have been
discussed in several works~\cite{gsv,admR} for the case  of asymmetric dark matter.
Typically it is seen that the effect of including the
asymmetry in the Boltzmann equations lead to a significant effect on the relic density. In these
works the analysis was done in the approximation  $\langle \sigma v \rangle = a+ b v^2$. Our analysis
below differs from these in that for our case annihilation via the $Z'$ pole is the dominant
process. Thus in our analysis we need to carry out an explicit thermal averaging over the
Breit-Wigner pole. It is convenient to work with the Boltzmann equations for the
quantities $f_{\psi}\equiv n_{\psi}/(h T^3)$, and $f_{\bar\psi}\equiv n_{\bar\psi}/(h T^3)$
where $n_{\psi}$  ($n_{\bar \psi}$) is the number density of particle $\psi$ ($\bar \psi$) and
the combination $hT^3$ appears in
the entropy per unit  volume, i.e.,  $ s=  (2\pi^2/45) hT^3$ where $h$ is the entropy degrees of freedom.
The Boltzmann equations obeyed by
$f_{\psi}$ and  $f_{\bar\psi}$  take the form
\begin{align}
  \frac{\d f_{\psi}}{\d x} & = \alpha \langle \sigma v\rangle (f_{\psi} f_{\bar \psi} - f^{\rm eq}_{\psi} f^{\rm eq}_{\bar \psi})\,,
\label{rel0}\\
  \frac{\d f_{\bar\psi}}{\d x} & = \alpha \langle \sigma v \rangle (f_{\psi} f_{\bar \psi} - f^{\rm eq}_{\psi} f^{\rm eq}_{\bar \psi})\,,
\label{rel1}
\end{align}
where $x=k_BT/m_{\psi}$ in which $k_B$ is the Boltzmann constant and hereafter we set $k_B=1$, and
$\alpha$ is given by
\beqn
\alpha(T) = \sqrt{90} m_{\psi} M_{\rm Pl} \frac{h}{\sqrt {g}\pi} \left(1+ \frac{1}{4}\frac{T}{g} \frac{\d g}{\d T}\right)\,,
\eeqn
where $g$ is the degrees of freedom that enter in the energy per unit volume, i.e.,
$\rho= \frac{\pi^2}{30} g T^4$, where $T(t)= T_{\gamma}(t)$ is the photon temperature.
Numerically $\alpha (T) = 6.7 \times 10^{20}$ GeV$^2$ for $g=h=68$ at $T=0.5$ GeV.
$\langle \sigma v \rangle$ is the thermally averaged cross section
\beqn
\langle\sigma v\rangle = \frac{\displaystyle{\int_0^{\infty}} \d v\,  (\sigma v) v^2 e^{-v^2/4x}}
{\displaystyle{\int_0^{\infty}} \d v\, v^2  e^{-v^2/4x}}\,.
\eeqn
Further, in Eqs.~(\ref{rel0}) and (\ref{rel1}) $f^{\rm eq}_{\psi}$ and  $f^{\rm eq}_{\bar \psi}$
are values of $f_{\psi}$ and $f_{\bar \psi}$ at equilibrium.
Now one can obtain the result from Eqs.~(\ref{rel0}) and (\ref{rel1}) that
the difference of $f_{\psi}$ and $f_{\bar \psi}$, i.e.,
\beqn
\gamma \equiv  f_{\psi} - f_{\bar \psi}\,,
\label{rel4}
\eeqn
is a constant. Assuming that the asymmetric dark matter currently constitutes a fraction $\lambda$ of the dark matter relic density,
one can evaluate $\gamma$ to be
\beqn
\gamma \simeq
\lambda \,\frac{5 \rho_c}{6\hbar T^3 m_{\psi}} \equiv \lambda \gamma_0, \quad \gamma_0 \approx 1.3\times 10^{-10}
\quad (m_{\psi} \sim 10\gev)\,,\label{Scale}
\eeqn
where the $5/6$ in $\gamma_0$ is due to Eq.~\eqref{ratio}.
\\

It is now straightforward to obtain the individual
relic densities for $\psi$ and $\bar \psi$.
Thus one integrates Eqs.~(\ref{rel0}) and (\ref{rel1}) from the freeze-out temperature to the current temperature
of $T_{\gamma}^0= 2.73$~K.  In the integration we will make the following approximation which
is conventionally done, i.e.,
we move $\alpha$ out of
the integral and replace it with $\alpha(x_f)$, i.e., by the value of $\alpha$ at the freeze-out temperature.
The matter density of $\psi$ at current temperature is given by $\rho_{\psi} = m_{\psi} n_{\psi}(x_0)$
where $x_0 =T_{\gamma}^0\big/m_{\psi}$ and $T_{\gamma}^0$ is the current photon temperature
of  $2.73$~K. The relic density  then is
\beqn
\Omega_{\psi} = m_{\psi} n_{\psi}(x_0) \big/ \rho_c\,,
\eeqn
where $\rho_c$ is the critical matter density so that $\rho_c= (3\times 10^{-12}\gev)^4 h_0^2$ where
$h_0$ is the Hubble parameter.  The integration of Eq.~(\ref{rel0}) straightforwardly gives
\beqn
\Omega_{\psi} h_0^2 = 2.2 \times 10^{-11} \sqrt{g(x_f)} h(x_0,x_f)
\left(\frac{T_{\gamma}}{2.73}\right)^3
\left (\frac{1}{\xi} - \frac{f_{\bar \psi}(x_f)}{ \xi  f_{\psi}(x_f)} e^{- \xi J(x_f)}\right)^{-1}\,,\label{Relic1}
\eeqn
where
\begin{align}
J(x_f)  \equiv \int_{x_0}^{x_f} \langle \sigma v \rangle \,\d x\,,
\qquad
h(x_0, x_f)  \equiv  \frac {h(x_0)}{h(x_f)}\left[1+ \frac{1}{4}\left(\frac{T}{g} \frac{\d g}{\d T}\right)_{x_f} \right]^{-1}\,,
\end{align}
and $\xi \equiv \alpha(x_f) \gamma$ where $\alpha(x_f)$ is the value of $\alpha$ evaluated at the
freeze-out temperature, and where
$g(x_f)$ ($h(x_f)$) are the energy (entropy) degrees of freedom at freeze out and $h(x_0)$ is the
entropy degrees of freedom at the current temperature.
The derivative term $\frac{1}{4}(\frac{T}{g} \frac{\d g}{\d T})_{x_f}$ is small and is often dropped,
while $h(x_0)= 3.91$~\cite{kt,Olive:1980wz}  and we estimate $h(x_f)\sim g(x_f)$ given $T_f$. As discussed below
$x_f$ is typically of size $\sim 1/20$ and thus $T_f=m_{\psi} x_f \sim 0.5\gev$ for $m_{\psi}\sim 10\gev$.
Now for $T_f \sim 0.5\gev$, $h(x_f)\sim 68$ which gives
$h(x_0,x_f)\sim 1/17.5$.
The quantities $f_{\psi}(x_f)$ ($f_{\bar \psi}(x_f)$) are $f_{\psi}$ ($f_{\bar \psi}$)
evaluated at freeze out.  Analogous to the relic density for $\psi$, we can get the relic density of
 $\bar \psi$ by integration of $\bar \psi$ and we obtain
\beqn
\Omega_{\bar \psi} h_0^2 = 2.2 \times 10^{-11} \sqrt{g(x_f)}
h(x_0,x_f)
\left(\frac{T_{\gamma}}{2.73}\right)^3
\left (\frac{f_{\psi}(x_f)}{ \xi  f_{\bar \psi}(x_f)} e^{\xi J(x_f)} - \frac{1}{\xi}\right)^{-1}\,.\label{Relic2}
\eeqn
The total dark matter relic density is
\beqn
\Omega_{\rm DM}= \Omega_{\psi} + \Omega_{\bar\psi}\,.
\eeqn
From Eqs.~(\ref{Relic1}) and (\ref{Relic2}) one obtains the ratio of the current relic densities
of $\bar \psi$ and $\psi$ to be
\beqn
\frac{\Omega_{\bar \psi} h_0^2}{\Omega_{\psi} h_0^2}= \frac{f_{\bar \psi}(x_f)}{f_{\psi}(x_f)} e^{- \xi J(x_f)}\,.
\label{omegaratio}
\eeqn
The front  factor ${f_{\bar \psi}(x_f)}/{f_{\psi}(x_f)}$  in Eq.~(\ref{omegaratio})
takes into account the asymmetry that exists at the freeze-out temperature. The size of this effect is estimated at the end of this section and could be as much as 20\%, and thus significant.
Our result of Eq.~(\ref{omegaratio}) is in agreement with the analysis of~\cite{gsv}.
\\

We discuss now the evaluation of the freeze-out temperature. The definition of this quantity
differs somewhat in various works (see, for example, \cite{Lee:1977ua,Scherrer:1985zt})
but these differences are not very significant. We adopt here the
definition of \cite{Lee:1977ua} where the freeze-out temperature $T_f$ is defined
as the temperature where the annihilation rate per unit volume equals the rate of change of the
number density. For our case this implies
\begin{align}
\frac{\d f^{\rm eq}_{\bar\psi}}{\d x}  = \alpha \langle \sigma v \rangle  f^{\rm eq}_{\psi} f^{\rm eq}_{\bar \psi}\,,
\quad{\rm at}~x= x_f = T_f/m_{\psi}\,,   \label{rel21}
\end{align}
where
$f^{\rm eq}_{\bar\psi}$ takes the form
\beqn
f^{\rm eq}_{\bar\psi}(x)= a_{\bar\psi}\, x^{-3/2} e^{-1/x}\,,
\label{rel23}
\eeqn
where $a_{\bar \psi} = g_{\bar \psi}(2\pi)^{-3/2}h^{-1}(T) \approx 9.3\times 10^{-4} g_{\bar \psi}$ around $T=0.5\gev$,
and $g_{\bar \psi}$ denotes the degrees of freedom of the dark particle
($g_{\psi}=g_{\bar \psi}=4$ for Dirac spinors).
The freeze-out temperature is then determined by the relation
\beqn
(x_f^{-1/2}- \tfrac{3}{2} x_f^{1/2})\, e^{1/x_f} = \alpha \langle \sigma v \rangle (
a_{\bar\psi} +  \gamma x_f^{3/2} e^{1/x_f})\,.
\label{rel24}
\eeqn
For the case of no asymmetry, i.e., in the limit  $\gamma \to 0$, Eq.~(\ref{rel24}) reduces down
to the well-known result~\cite{Lee:1977ua}. {One may compare the analysis of the freeze-out
temperature given by Eq.~\eqref{rel24} with the one
using the alternate criterion~\cite{Scherrer:1985zt}
\beqn
\Delta(x_f) = c f^{\rm eq}_{\bar\psi}(x_f)\,,
\label{rel25}
\eeqn
where
$\Delta (x) \equiv (f_{\bar \psi} (x)- f^{\rm eq}_{\bar \psi}(x))$
and $c$ is order unity. Using Eq.~\eqref{rel23} in Eq.~\eqref{rel25} one gets
\beqn
(x_f^{1/2}- \tfrac{3}{2} x_f^{-1/2} -\alpha \gamma \langle \sigma v \rangle x^{3/2})\, e^{1/x_f}
  = \alpha a_{\bar\psi} c(c+2) \langle \sigma v \rangle \,.
\label{rel26}
\eeqn
For $\gamma=0$, Eq.~(\ref{rel26}) reduces to the result of \cite{Scherrer:1985zt}
while $\gamma\neq 0$ gives the correction due to asymmetry.
Further, we see that Eq.~(\ref{rel26}) reduces to Eq.~(\ref{rel24}) when $c=\sqrt 2 -1$.
To compute the sensitivity of the freeze-out temperature on the asymmetry it is useful
to utilize the scale factor $\lambda$ defined in Eq.~\eqref{Scale}.
On using Eq.~(\ref{rel24}) we can obtain an approximate expression for $\d x_f/\d \lambda$ so that
\beqn
\d x_f/\d \lambda \simeq - a_{\bar \psi}^{-1} \gamma_0 x_f^{7/2} e^{1/x_f}\,.
\eeqn
From above we can compute the first order correction to the freeze-out temperature
due to the asymmetry. To the leading order one has
\beqn
x_f \simeq x_f^0 \left[1- a_{\bar \psi}^{-1} \gamma (x_f^0)^{5/2} e^{1/x_f^0}\right]\,,
\eeqn
where $x^0_f$ is the zeroth order of the $x_f$, i.e., when $\gamma =0$.
We note that the correction to the freeze-out temperature due to asymmetry
is independent of $\langle \sigma v \rangle$ to leading order.
Using $a_{\psi}= 3.7\times 10^{-3}$, $x_f=1/17.5$ and
$\gamma = \gamma_0 = 1.3\times 10^{-10}$,
one finds that the correction to $x_f$ is around a percent
for the choice of the parameters given.
Further, as $\gamma ~({\rm and~hence} ~ \xi) \to 0$, one has
$\frac{f_{\psi}(x_f)}{f_{\bar \psi}(x_f)}\to 1$ and in this limit one has
\beqn
\Omega_{\psi} h_0^2 = \Omega_{\bar\psi} h_0^2
= 2.2 \times 10^{-11} \sqrt{g(x_f)} h(x_0,x_f)
 \left(\frac{T_{\gamma}}{2.73}\right)^3 \frac{1}{J(x_f)}\,.
\eeqn
\\

Now rapid annihilation of dark matter can
occur if the sum of the dark matter masses is close to the
$\zp$ pole  and there is a Breit-Wigner enhancement~\cite{Griest:1990kh,Nath:1992ty}.
Thus for the case we are considering if the mass of the $\zp$
is close to twice the mass of the dark particle, then
one can get a large annihilation cross section and  correspondingly
a small relic density.
An analysis of the relic density arising from the annihilation of symmetric dark matter is given
in  Fig.~\ref{rel_den} and the analysis shows that   the
relic density arising from the symmetric component of dark matter can easily be
made negligible, i.e., less than $10\%$ of the cold dark matter
density given by WMAP.
In Fig.~\ref{rel_den}  we give the analysis for the case with no asymmetry, i.e., $\gamma=0$
(left panel) and the  case with asymmetry (right panel) where
 $\gamma=1.3\times10^{-10}$.  A comparison of the left and the right panels shows that inclusion of the
 asymmetry has a substantial effect on the relic density. Specifically it further helps deplete the
 relic density of $\bar\psi$ (the symmetric component of dark matter).
 For the case of $g_C=1$ the allowed upper bound of the $\zp$ mass
 increases by about $\sim 100\GeV$ in the presence of an asymmetry when $\gamma =1.3\times 10^{-10}$.
It is also instructive to examine the ratio of the thermal relic density for the  cases with and without asymmetry.
Here one has
\beqn
{\rm R}\,\equiv \frac{(\Omega_{\bar{\psi}}h_0^2)_{\gamma=\gamma_0}}
{(\Omega_{\bar{\psi}}h_0^2)_{\gamma=0}}=\frac{\xi J\left(x_f\right)}{\frac{f_{\psi}(x_{f})}{f_{\bar{\psi}}(x_f)} e^{\xi J\left(x_f\right)}-1} \,.
\label{rel_r}
\eeqn
As $\xi \to 0$,
$\frac{f_{\psi}(x_{f})}{f_{\bar{\psi}}(x_f)}\to 1$ and thus $R\to 1$. However, if we assume that
the asymmetric dark matter is responsible for $5/6$ of the total relic density, then for
$m_{\psi}\sim 10\GeV$, one has $\gamma=1.3\times 10^{-10}$ and $f_{\bar\psi}(x_f)=6.8\times 10^{-10}$
which gives  $\frac{f_{\psi}(x_{f})}{f_{\bar{\psi}}(x_f)} = (1+ \gamma_0/f_{\bar{\psi}}(x_f))
\sim 1.2$. In this circumstance one finds that
$R$ is always less than 1. Thus one finds that the inclusion of asymmetry helps deplete the
symmetric component of dark matter.

\begin{figure}[t!]
\begin{center}
\includegraphics[scale=0.23]{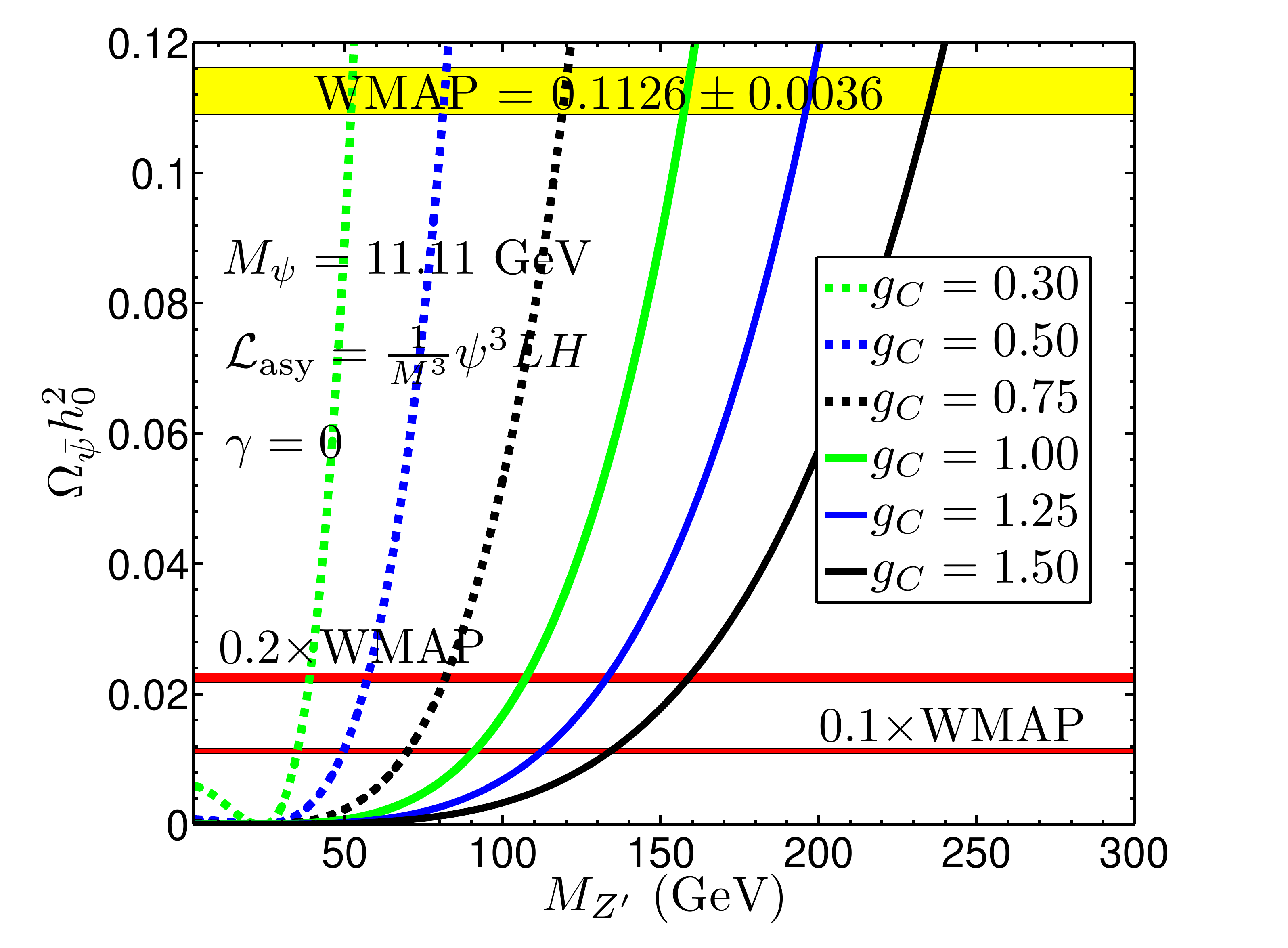}
\includegraphics[scale=0.23]{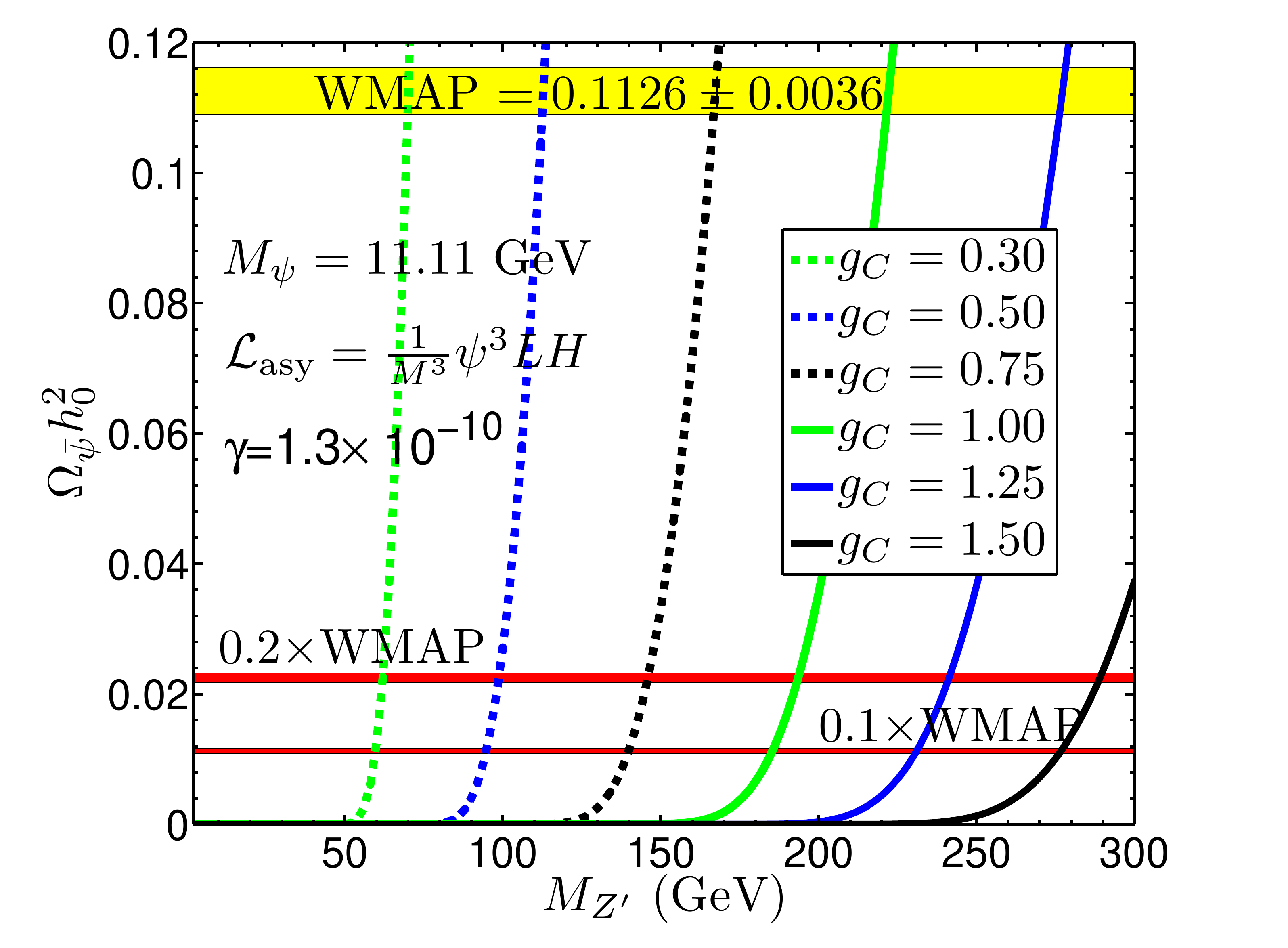}
\caption{\label{rel_den} (color online)
An exhibition of the thermal relic density of $\bar{\psi}$ as a function of $M_{\zp}$
in the model with gauged $L_{\mu}-L_{\tau}$ for different values of the coupling constant.
The left panel shows the case $\gamma=0$ and the right panel shows $\gamma=1.3\times 10^{-10}$.
In both cases, the analysis shows that the component of dark matter that is thermally produced can be efficiently
depleted by resonant annihilation via the $Z'$ pole.}
\end{center}
\end{figure}

\section{Asymmetric dark matter in a \st extension of the MSSM} \label{Sec:ASY_MSSM}
The analysis of dark matter in the MSSM extension is more complex in that there are now three
contributions to the dark matter relic density, i.e., from the asymmetric and symmetric components
as in Eq.~(\ref{6.1}) and from the neutralino. Thus here one has
\beqn
\Omega_{\rm DM} = \Omega_{\rm DM}^{\rm asy} + \Omega_{\rm DM}^{\rm sym}  + \Omega_{\tilde\chi^0}\,,
\eeqn
where $\Omega_{\tilde \chi^0}$ is the relic density from the neutralino.
In this case  for the asymmetric dark matter to work, one must significantly
deplete not only the symmetric component of dark matter but also the contribution from
the neutralino. Thus here we take the criterion that
$\Omega_{\rm DM}^{\rm sym}\,/\,\Omega_{\rm DM} <0.1$, and
$\Omega_{\tilde \chi^0}\,/\,\Omega_{\rm DM} <0.1$.
For the analysis of AsyDM in extensions of MSSM we consider the interaction
\begin{equation}
W_{{\rm asy}}=\frac{1}{M^3_{{\rm asy}}}X^{2} (LH_u)^2\,.
\label{susytransfer}
\end{equation}
Here we note that the choice $W_{\rm asy} \sim X^{2}LH_u$ would have allowed the decay
$\no \rightarrow XX\nu \cdots $ and would have required the constraint $m_{\no} < 2 m_{X}$ for the neutralino
to be stable. Further, while  the choices $W_{\rm asy} \sim X^{2}LLe^c, X^{2}Lqd^c$
do not allow the neutralino decay at the tree-level, such a decay can occur
at the loop level since it is not forbidden by a symmetry.
Additionally  $W_{{\rm asy}} \sim X^{3}LH_u, X^{3}LLe^c, X^{3}Lqd^c$ can also preserve the stability
of the neutralino. Here and elsewhere we are assuming that the
Stueckelberg neutralinos are heavier than the lightest neutralino in the MSSM sector
(see the discussion following Eq.~(\ref{e5})).
Returning to Eq.~\eqref{susytransfer},
the corresponding dark particle masses are computed to be
5.55~GeV (Model E) and 3.25~GeV (Model F).
Now the \st extension of MSSM, is more complex than the SM extension.
We exhibit the relevant parts of this extension below.
\\

For the \st Lagrangian of the supersymmetric case we choose\cite{kn12}
\beqn
\Lag_{\rm St} = \int d\theta^2 d\bar{\theta}^2\, \left[ MC+ S +\bar S \right]^2\,,
\label{mass}
\eeqn
where $C$ is the $U(1)_C$ vector multiplet, $S$ and $\bar S$ are chiral multiplets, and
$M$ is a mass parameter.
We define $C$ in the Wess-Zumino gauge as
$C= -\theta\sigma^{\mu}\bar \theta C_{\mu}
+i\theta\theta \bar\theta \bar \lambda_C
-i\bar\theta\bar\theta \theta  \lambda_C
+\frac{1}{2} \theta \theta\bar\theta\bar\theta D_C\,$,
while
$S=\frac{1}{2}(\rho +ia ) + \theta\chi
 + i \theta\sigma^{\mu}\bar\theta \frac{1}{2}(\partial_{\mu} \rho
+i \partial_{\mu} a)
+~ \theta\theta F +\frac{i}{2} \theta \theta \bar\theta \bar\sigma^{\mu} \partial_{\mu} \chi
+\frac{1}{8}\theta\theta\bar\theta\bar\theta (\Box \rho+i\Box a)\,$.
Its complex scalar component contains the axionic pseudo-scalar $a$, which is the analogue of the
real pseudo-scalar that appears in the non-supersymmetric version in \cite{kn12}.
We write $\Lag_{\rm St}$ in component
notation as (see e.g. \cite{kn12})
\beqn \label{stueck}
\Lag_{\rm St} = - \frac{1}{2}(MC_{\mu} +\partial_{\mu} a)^2
- \frac{1}{2} (\partial_\mu \rho)^2
- i \chi \sigma^{\mu} \partial_{\mu}\bar {\chi} + 2|F|^2
+ M \rho D_C
+ M\bar {\chi} \bar \lambda_C
 +  M\chi \lambda_C  \ .
\label{ls}
\eeqn
For the gauge fields we add the kinetic terms
\beqn
\Lag_{\rm gkin} =
-\frac{1}{4}   C_{\mu\nu} C^{\mu\nu}
- i \lambda_C \sigma^{\mu}\partial_{\mu} \bar\lambda_C
+\frac{1}{2} D_C^2\,,
\eeqn
with $C_{\mu\nu}\equiv\partial_\mu C_\nu - \partial_\nu C_\mu$.
For the matter fields (quarks, leptons, Higgs scalars, plus hidden sector matter)
chiral superfields  with components $(f_i,z_i,F_i)$ are introduced and the matter Lagrangian
is given by
\beqn \label{matt}
\Lag_{\rm matt} =
- |D_\mu z_i|^2 - i f_i \sigma^\mu D_\mu \bar f_i
- \left(
 i  \sqrt 2 g_C Q_C z_i \bar f_i \bar \lambda_C
                   +\, {\rm h.c.}\, \right)
+ g_C D_C (\bar z_i Q_C z_i) + |F_i|^2
\ ,
\eeqn
where  $(Q_C,g_C)$ are the charge operator and coupling constant
of  $U(1)_C$, and
$D_\mu = \partial_\mu +  ig_C Q_C C_\mu$ is the gauge covariant derivative.
It is  convenient to replace the two-component Weyl-spinors $(\chi,\bar{\chi}),(\lambda_C,\bar{\lambda}_C)$ by
four-component Majorana spinors,
which we label as
$\psi_S = (\chi_\alpha , \bar\chi^{\dot\alpha})^T$, and
$\lambda_C = (\lambda_{C\alpha},\bar\lambda_C^{\dot\alpha})$.
The total Lagrangian of the MSSM then takes the form
\beqn  \label{delta}
\Lag_{\rm StMSSM} &=& \Lag_{\rm MSSM} +
\Lag_{U(1)}+
\Delta \Lag_{\rm St}~,
\eeqn
with
\beqn
\Delta \Lag_{\rm St} &=& -\frac{1}{2}(MC_{\mu} +\partial_{\mu} a)^2
-\frac{1}{2} (\partial_\mu \rho)^2
- \frac12 M^2  \rho^2 \non
&&
- \frac{i}{2} \bar\psi_S \gamma^{\mu} \partial_{\mu} \psi_S
- \frac14 C_{\mu\nu}C^{\mu\nu}
- \frac{i}{2} \bar\lambda_C \gamma^\mu \partial_\mu \lambda_C
+ M \bar \psi_S\lambda_C  \non
&&
- \sum_i \Big[ \left| D_\mu z_i \right|^2 - \left| D_\mu z_i \right|^2_{C_\mu=0}
+ \rho   g_C M ( \bar z_i Q_C z_i )  \non
&&
\hspace{0.0 cm}
+ \frac12 g_C C_\mu \bar f_i \gamma^\mu Q_C f_i
+ \sqrt 2 g_C \Big( i z_i Q_C \bar f_i \lambda_C +\, {\rm h.c.}\, \Big) \Big]
  - \frac12 \Big[g_C \sum_i \bar z_i Q_C z_i \Big]^2\,.
  \label{4.j}
\eeqn
As in the SM case we assume that the $U(1)_C$ is a gauged $L_{\mu}- L_{\tau}$.
Further, we assume that all hidden sector fields while charged under $U(1)_C$ are  neutral
under the MSSM gauge group and some of the MSSM particles, i.e., the second and the third
generation leptons, are charged under $U(1)_C$.
As discussed already an essential ingredient to explain the cosmic coincidence is that the symmetric
component of dark matter produced in thermal processes is significantly depleted.  For the MSSM \st extension
the analysis of annihilation is essentially identical to the case of the \st extension of the standard model
and we do not discuss it further.
\\

We now discuss the fate of the extra particles that arise in the $U(1)_C$ \st extension of MSSM.
This extension involves the following set of particles:  $Z', \rho, \psi,\phi,  \psi_S, \lambda_C$.
The decay of the $Z'$ has already been discussed. Next we consider the $\rho$.
Eq.~(\ref{4.j}) gives the interaction of the  $\rho$ with the sfermions. Specifically its couplings  to the mass diagonal
sfermions  are given by
\beqn
\Lag_{\rho\tilde f ^{\dagger} \tilde f} & =& -g_{\rho} M_\rho \left[
\cos(2\theta_{\tilde f_i}) \left(\tilde f^{\dagger}_{1i} \tilde f_{1i}  -\tilde f^{\dagger}_{2i} \tilde f_{2i}
\right) +
\sin(2\theta_{\tilde f_i}) \left(\tilde f^{\dagger}_{1i} \tilde f_{2i}  +\tilde f^{\dagger}_{2i} \tilde f_{1i}
\right)  \right]\,,
\label{4.k}
\eeqn
where $f_i$ refer to $\mu, \nu_{\mu}, \tau,\nu_{\tau}$. Thus the $\rho$ will decay via second and third generation
slepton loops into $\mu^+\mu^-, \nu_{\mu}\bar \nu_{\mu}$, $\tau^+\tau^-$,  $\nu_{\tau}, \bar \nu_{\tau}$
and disappear in the thermal bath quickly  (see Appendix~\ref{App:rhoDR}).
Next we  discuss the neutralino sector. Here in the  $U(1)_C $  \st extension of MSSM the neutralino sector is enlarged  in that one has two more fields, i.e., the  gaugino, and the higgsino fields $(\Psi_S, \Lambda_C)$ as mentioned earlier. In this case the neutralino mass matrix of the $U(1)_C$  extension of MSSM is given by
 \beqn
 \cal{M}_{\rm neutralino} =
\left(
\begin{array}{c|c}
{\cal M}_{st} &  0_{2 \times 4} \\
 \hline
 0_{4 \times 2} &  {\cal M}_{\rm MSSM} \\
\end{array}
\right)\,,
\qquad
{\cal{M}}_{\rm st}  =
   \left( \begin{array}{cc}
    0 & M  \\
    M & \tilde M  \\
 \end{array} \right)\,,
 \label{e5}
\eeqn
where $M_{\rm St}$ is in the basis $(\Psi_S, \Lambda_C)$, $M$ is the \st mass and  $\tilde M$ is the soft mass.
The neutralino mass eigenstates arising from  Eq.~\eqref{e5} can be labeled $\nsta,\nstb$.
We consider the possibility that the \st neutralinos are heavier than the LSP of the MSSM ($\na$) and decay into the MSSM
neutralino which is assumed to be stable. In this case one will have more than one dark
matter particle, i.e.,  the $\psi$ from the \st sector and $\na$ from the MSSM sector.
Again in the case of AsyDM the relic density of $\na$ must be much
smaller than the WMAP relic density for CDM. To this end
we carry out an explicit analysis of the relic density within supergravity~(SUGRA) grand unification~\cite{can}.
As we show in Fig.~\ref{fig_sugra} the relic density of $\na$
 can be very small (see the next section for more detail),
which allows the dominant component of the dark matter observed today to be the asymmetric dark matter.

\begin{figure}[t!]
\begin{center}
\includegraphics[scale=0.2]{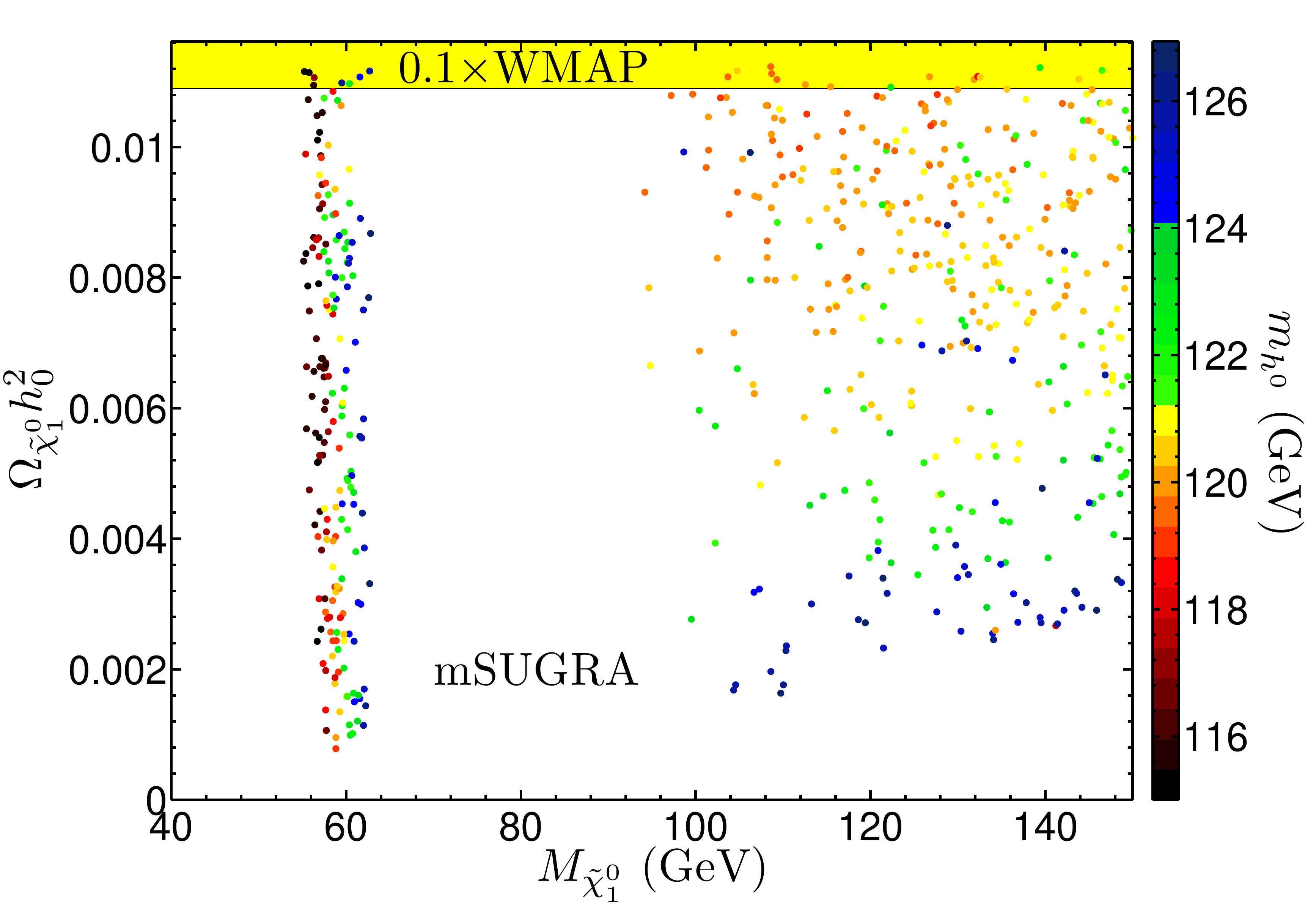}
\includegraphics[scale=0.2]{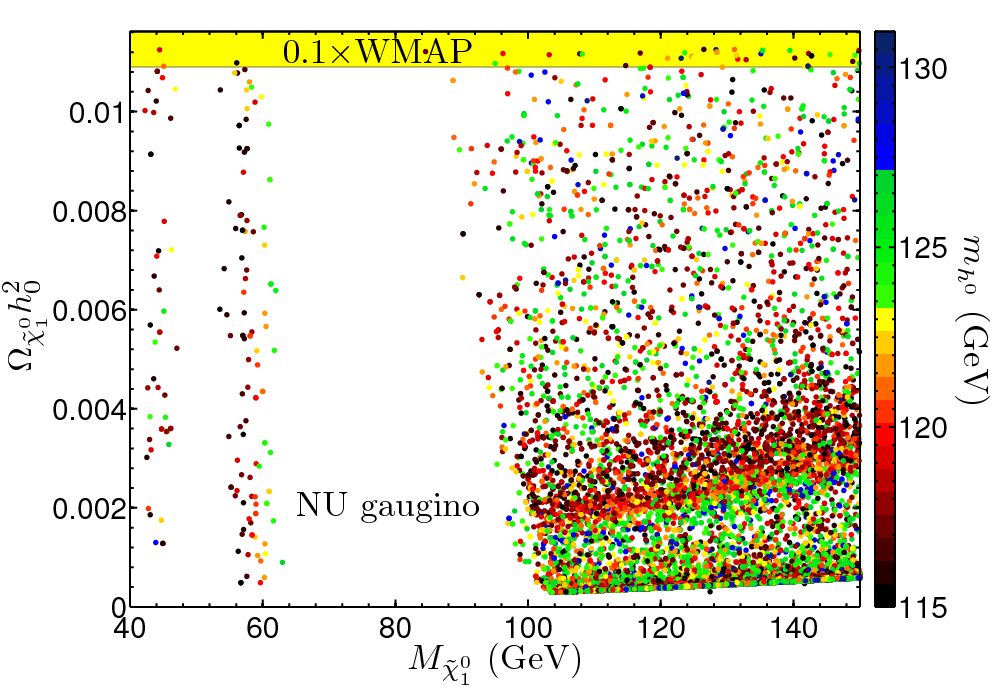}
\includegraphics[scale=0.2]{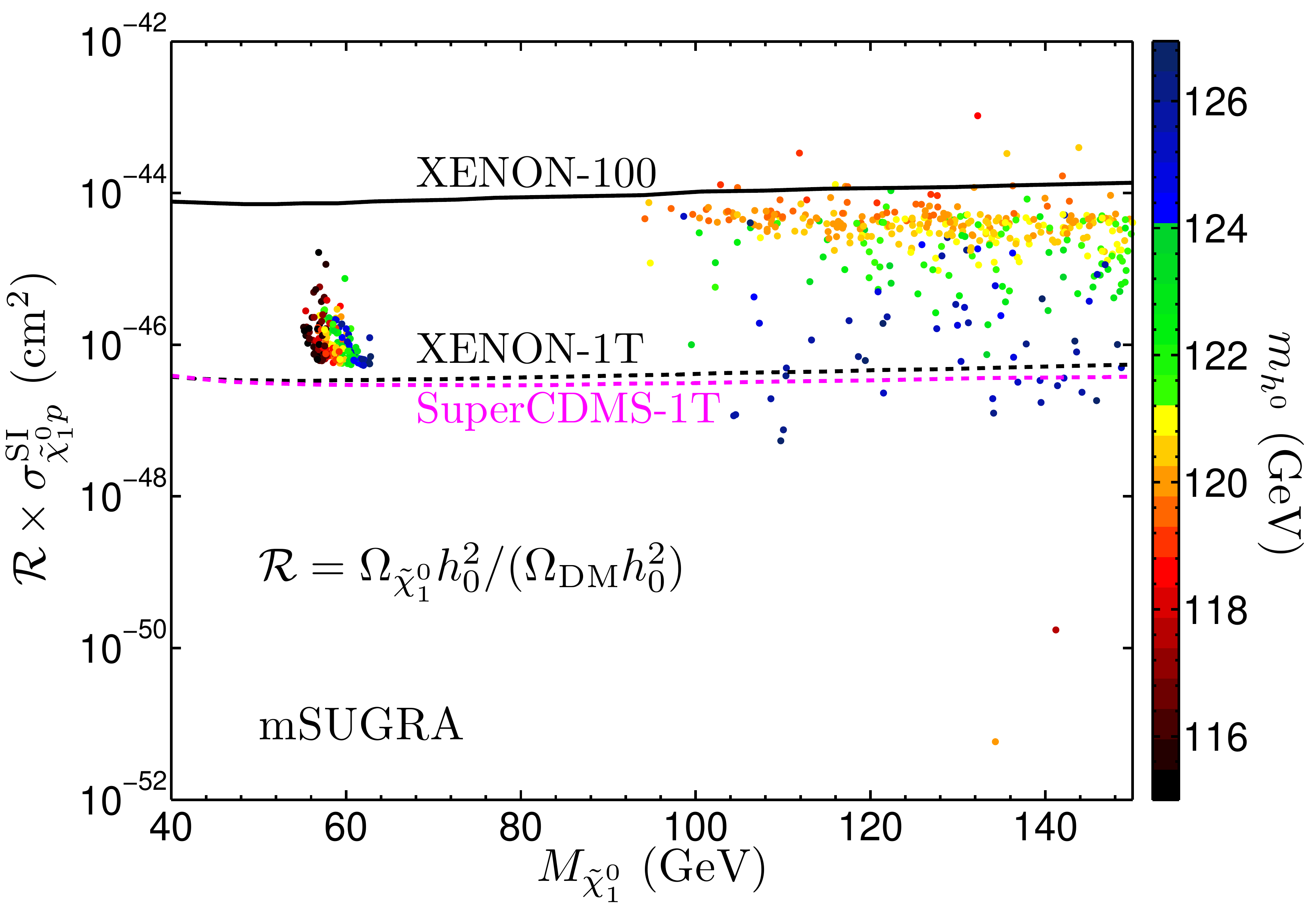}
\includegraphics[scale=0.2]{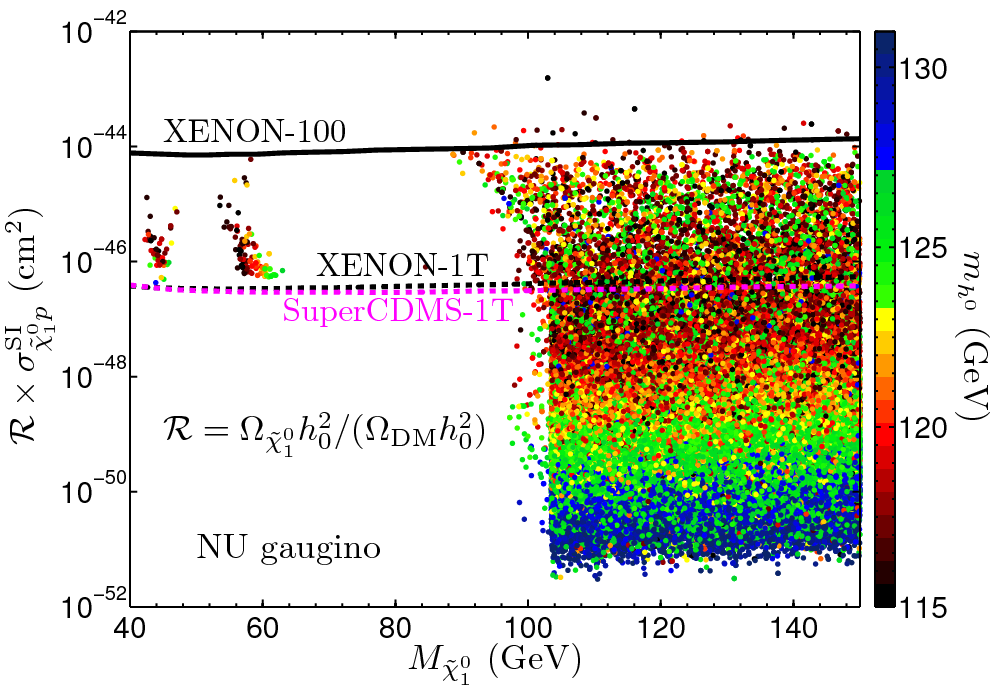}
\caption{\label{fig_sugra}
(color online)
Top panel:
An exhibition of the depletion of the MSSM neutralino dark matter below $10\%$  of the WMAP relic
density for cold dark matter.
Parameter points are displayed by their light CP even Higgs mass and the yellow band
corresponds to $10\%$ of the WMAP-7 observed limit.
Bottom panel: An exhibition of the neutralino-proton spin-independent cross section  as a function of the neutralino mass.
To account for the reduced relic density of the neutralino component of  dark matter
the spin-independent cross section has been corrected by a factor
 $\mathcal{R}=\Omega_{\na} h_0^2 / \left(\Omega_{\rm DM} h_0^2\right)$.
The present experimental limits (solid line)~\cite{xenon} as well as the future projections (dashed lines)
 are shown~\cite{futureXENON,futureCDMS}.
The left panel shows the parameter points of mSUGRA
and the right panel shows the non-universal gaugino parameter points.
All parameter points shown pass the general constraints.
}
\end{center}
\end{figure}

\section{Detection of dark matter} \label{Sec:Ex_De}
The AsyDM in the model we consider can interact with the standard model particles
only via the $Z'$ and $\rho$ bosons which couple with the
second and third generation leptons. Thus the scattering of AsyDM from nuclear targets
will not produce any visible signals and the detection of AsyDM in direct detection experiments
is difficult.
However, it is interesting to investigate if the subdominant component of dark matter could
still provide a detectable signature.  We discuss this topic in further detail below.
First we discuss the  depletion of $\na$ dark matter to determine
the regions of the parameter space where the relic density of $\na$ is a negligible fraction
of the WMAP relic density for CDM, and is thus indeed a subdominant component of dark matter.
Later we will investigate the possibility of detection of this subdominant component in
direct detection experiments.
Specifically we investigate two classes of models:
the supergravity grand unified model (mSUGRA) with universal boundary conditions on soft parameters at the GUT scale,
and non-universal SUGRA model with non-universalities in the gaugino sector (see, e.g.,
\cite{Chattopadhyay:2001mj} and the references there in).
\\

For the mSUGRA case the parameter space investigated was:
 $m_0<10\TeV$, $\mhf<10\TeV$, $\left|A_0/m_0\right|<10$,
$1<\tan\beta<60$, and $\mu>0$.
For the case of SUGRA models with non-universalities in
the gaugino sector the parameter space  investigated was:
$M_i=\mhf\left(1+\delta_i\right)$ with the same
ranges as in the mSUGRA case with $\left|\delta_i\right|<1 $.  After radiative breaking of the electroweak
symmetry we collected roughly $31.4$~million mSUGRA models and
$25.6$~million non-universal~(NU)  gaugino models.
These models were then subjected to the experimental constraints which included
the limits on sparticle masses from LEP~\cite{pdgrev}:
$m_{\sta} > 81.9\GeV$,
$m_{\cha} > 103.5\GeV$,
$m_{\ta} > 95.7\GeV$,
$m_{\ba} > 89\GeV$,
$m_{\ser} > 107\GeV$,
$m_{\smr} > 94\GeV$,
and $m_{\g} > 308\GeV$.
as well as the recent bounds on the light CP even, SM-like, Higgs
from ATLAS and CMS, i.e  $115\GeV < m_{h^0}< 131\GeV$~\cite{ATLAShiggs,CMShiggs}.
More recently, ATLAS has constrained the SM-like Higgs to be between $\left(117.5-118.5\right)\GeV$ and $\left(122.5-129\right)\GeV$ and CMS has  constrained the Higgs Mass to be between $\left(115-127\right)\GeV$~\cite{higgsUPdate}.
This constraint applied to the mSUGRA parameter space has recently been
discussed in~\cite{Akula:2011aa}.
As discussed above  if $\na<\nsta$ then the neutralino would contribute to the
relic density and for the AsyDM model to work we require that  $\Omega_{\na} h_0^2$, to be less then $10\%$
of the WMAP-7 limit~\cite{wmap}.
Other constraints applied to the parameter points include the $g_\mu -2$~\cite{gmuon2} constraint discussed in Section~\ref{Sec:ASY_SM} and constraints {from}
B-physics~{measurements}~\cite{bphys,cmslhcbbsmumu,Abazov:2010fs}
which yield flavor constraints, i.e.
$\left(2.77\times 10^{-4} \right)\leq {\mathcal{B}r}\left(b\to s\gamma\right) \leq \left( 4.37\times 10^{-4}\right)$
(where this branching ratio has the NNLO correction~\cite{Misiak:2006zs})
and
${\mathcal{B}r}\left(B_{s}\to \mu^{+}\mu^-\right)\leq 4.5\times10^{-9}$.
As done in~\cite{Akula:2011jx}, we will refer to these constraints as the  {\it general constraints}.  These constraints were
done by calculating the sparticle mass spectrum with {\sc SuSpect}~\cite{suspect}
and using {\sc micrOMEGAs}~\cite{belanger} for the  relic density as well as  for the indirect constraints.
\\

In Fig.~\ref{fig_sugra}, we exhibit the mSUGRA~(left panels) and the
NU gaugino~(right panels) parameter points after applying the general constraints.  In the top two panels we show the thermal
relic density of the neutralino. As discussed previously, one finds that there is a significant region of the
parameter space with a relic density much less than one tenth of the WMAP relic density.
Thus the neutralino is indeed a subdominant component of dark matter.
There are many more NU gaugino parameter points that satisfy the relic density compared
to mSUGRA parameter points.  This comes about because of coannihilation.
Thus the non-universal case allows for the light chargino to lie close to
the LSP, i.e. $(m_{\cha}-m_{\na})/m_{\na} \ll 1$, allowing for coannihilation to occur.
The relevant question then is
if such a subdominant component can be detected in dark matter experiments. This is exhibited
in the  lower two panels of Fig.~\ref{fig_sugra}, where
the corrected neutralino-proton spin-independent cross section, i.e
$\mathcal{R} \times\sigma_{\na p}^{\rm SI}$ where
$\mathcal{R}=\Omega_{\na} h_0^2 / \left(\Omega_{\rm DM} h_0^2\right)$, is given as a function of the neutralino mass. For comparison
we show the current XENON-100 bound~\cite{xenon} and the projected bounds in future
experiments~\cite{futureXENON,futureCDMS}.
The important observation is that  neutralino-proton spin-independent cross section is still detectable even
when the neutralino is a subdominant component of dark matter with a relic density less than $10\%$  of the WMAP relic density for CDM.

\begin{figure}
\begin{center}
\includegraphics[scale=0.8]{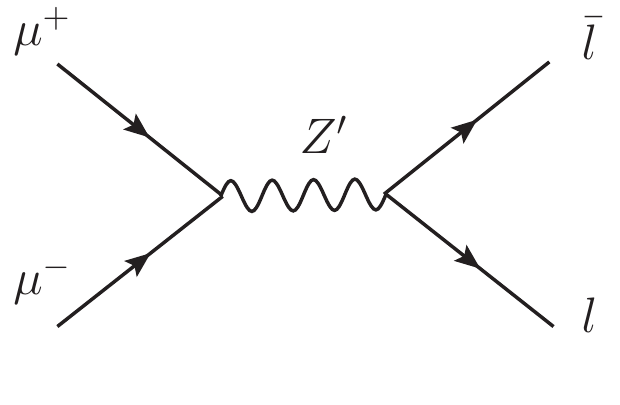}\qquad\qquad
\includegraphics[scale=0.8]{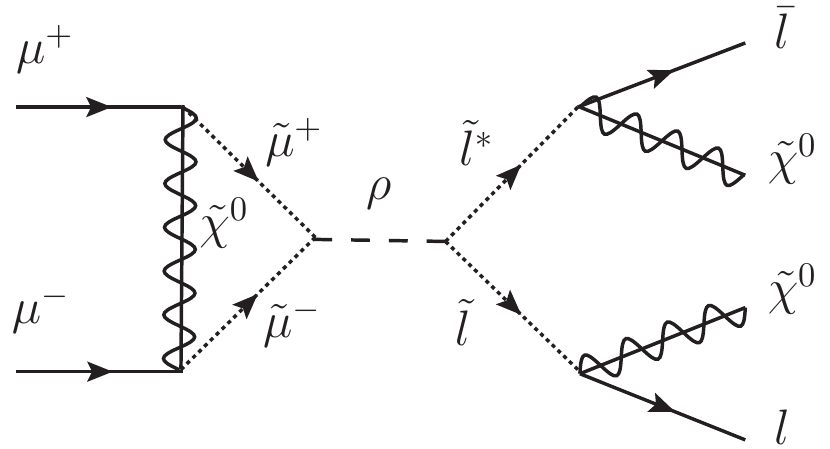}
\caption{\label{FD_mucol}
Left: Leptonic final states in a $\mu^+\mu^-$ collider where the $\mu^+\mu^- \to l\bar l$, with $l=\mu,\nu_{\mu},\tau,\nu_{\tau}$,  final state
arising from direct channel poles involving $\zp$.  The $Z'$ pole does not
allow for a  $e^+e^-$ final state and thus the relative production cross section for $\mu^+\mu^- \to
\tau^+\tau^-$ vs $\mu^+\mu^-\to e^+e^-$ can be used to detect the existence of a $L_{\mu}- L_{\tau}$ gauged boson. Right: A similar analysis is possible for $\rho$ but its production
is suppressed relative to $\zp$ since it must be produced at the loop level.}
\end{center}
\end{figure}
\begin{figure}
\begin{center}
\includegraphics[scale=0.27]{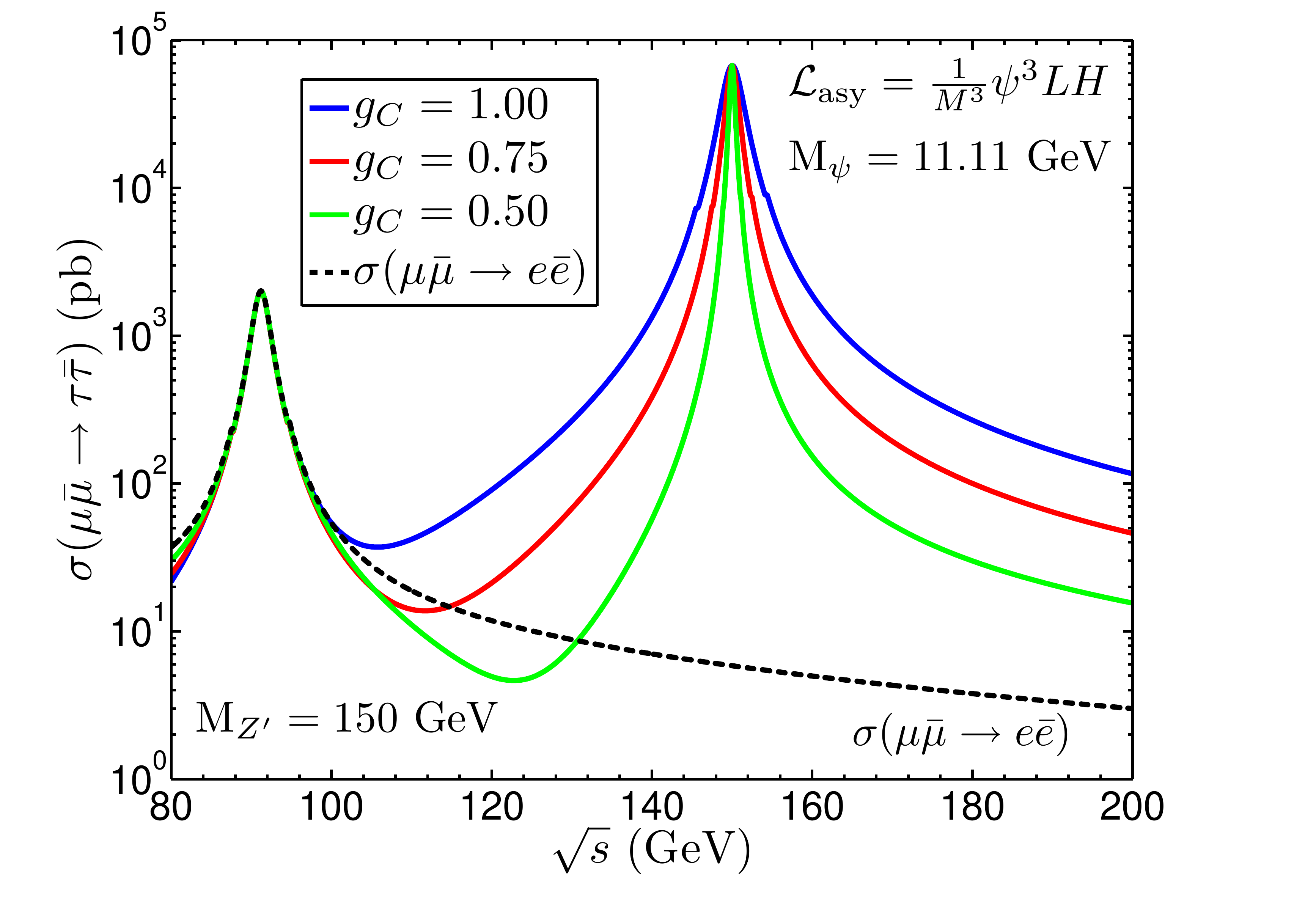}
\caption{\label{mucollider} (color online) An exhibition of the relative strength of the $\tau^+\tau^-$ vs
$e^+e^-$ signal at a muon collider. The presence of a detectable $Z'$ resonance in the $\mu^+\mu^-\to\tau^+\tau^-$ channel provides a smoking gun signature for the gauged
$L_{\mu}- L_{\tau}$ AsyDM model. A similar resonance is also present in the
$\mu^+\mu^-\to\mu^+\mu^-$ channel while $\mu^+\mu^-\to e^+e^-$ cross section shows
no such enhancement in the $\zp$ region.}
\end{center}
\end{figure}

\section{Signatures at colliders} \label{Sec:Ex_Co}
The AsyDM models discussed above can produce a dramatic signature at a muon collider, see Fig.~\ref{FD_mucol},
which we now discussed (Signatures of a $Z'$ boson in a gauged $U(1)_{L_\mu - L_\tau}$ model
at a muon collider were discussed in~\cite{Baek:2001kca} but the analysis was only at the
tree-level.). This signature arises from a $Z'$ resonance. We note that
$Z'$  has no couplings with the first generation leptons and thus  a process such as
$e^+e^-\to Z' \to \mu^+\mu^-, \tau^+\tau^-$ is absent at the tree-level. This process
can only arise at the loop level which, however, is suppressed relative to the tree. This
explains why such a resonance has not been observed yet at an $e^+e^-$ collider (see Appendix~\ref{App:Z'}).
However, dramatic signals will arise at a muon collider where  we will have processes of the type
\begin{gather}
\mu^+\mu^- \to  \zp \to \mu^+\mu^-, \nu_{\mu} \bar \nu_{\mu},\,\tau^+\tau^-, \nu_{\tau} \bar \nu_{\tau}\,. \nonumber
\end{gather}
Since the final states contain no $e^+e^-$ this would provide a smoking gun signature for the model.
In Fig.~\ref{mucollider} we exhibit the cross section $\sigma(\mu^+\mu^-\to \tau^+\tau^-)$ for various
values of $g_C$ when the AsyDM mass is taken to be 11.11~GeV and the $Z'$ mass is 150~GeV. For
comparison $\sigma(\mu^+\mu^-\to e^+e^-)$ is also plotted. One finds that the
$\sigma(\mu^+\mu^-\to  \tau^+\tau^-)$ exhibits a detectable  $Z'$ resonance and the cross section
varies dramatically  as a function of $\sqrt s$ relative to
$\sigma(\mu^+\mu^-\to  e^+e^-)$ which is a rather smoothly falling function beyond the Z boson pole.  In Appendix~\ref{App:Z'} it is shown that the loop contribution to
$\mu^+\mu^-\to e^+ e^-$ is suppressed and the $\zp$ resonance is not discernible in
this channel at a  $\mu^+\mu^-$ collider.
We note that there is a second overlapping  resonance from a spin 0 $\rho$ state where
$ \mu^+\mu^-\to \rho \to \tilde \mu^*\tilde \mu \to \mu^+\mu^-  2\tilde  \chi^0$. However, the
$\rho$ resonance can only proceed at the loop level and  is suppressed  relative to the
$Z'$ pole.

\section{A gauged $B-L$ model} \label{Sec:B-L}
Next we discuss briefly the case where in the \st extension we use $U(1)_{B-L}$
rather than $U(1)_{L_{\mu}-L_{\tau}}$.
Here we consider models \textit{with} the right-handed neutrinos in order to gauge $B-L$
and  focus on the model $\rm A_2'$ with the $B-L$ transfer interaction
$\mathcal L_{\rm asy} = \frac{1}{M^4}\psi^2\left(LH\right)^2$
above the EWPT scale. By including three generations of right-handed neutrinos,
the asymmetric dark matter mass is computed to be 6.06~GeV, c.f. Table~\ref{tab_3}.
In this case, there are more experimental constraints to consider
which include collider (i.e., LEP, Tevatron, LHC) constraints as well as
precision constraints (i.e., the measurements of the $\rho$ parameter, the
$\Upsilon$ width).  Specifically the LEP constraint gives
$M_{\zp}/g_C^{\prime}\gtrsim 6\TeV$~\cite{Carena:2004xs} for heavy gauge
bosons. A stricter bound within a specific framework is given in ~\cite{Kim:2011xv}
where $M_{Z'} \geq 10$~TeV.
For lighter gauge bosons, as in the case of~\cite{Liu:2011di,zpUA2},  the UA2
cross section bound~\cite{ua2} is more stringent.  Our analysis here is consistent
with these constraints.
Now, as in the $L_{\mu}-L_{\tau}$ case, the thermal symmetric contribution to the
relic density from AsyDM must still be consistent with WMAP, i.e. it must be depleted
to below $10\%$ of the WMAP-7 value.
An analysis of this is given in Fig.~\ref{fig_bl_rel} for the model $\rm A_2'$
with $\gamma=0$~(dashed line) and $\gamma=\gamma_0=1.3\times10^{-10}$~(solid line).
Here one finds that the symmetric component of AsyDM can satisfy our WMAP-7 constraint
for a range of $\zp$ masses. If one wishes to keep the $\zp$ mass in
the $\Upsilon$ region, i.e. $\sim 10\GeV$, then a fine-tuned value of $g^{\prime}_C$ is needed as
seen in Fig.~\ref{fig_bl_rel} to be consistent with the constraints on $M_{\zp}/g^{\prime}_C$.
Heavier $\zp$ masses would have difficulty satisfying both the collider and WMAP-7 constraints
discussed above.

\begin{figure}[t!]
\begin{center}
\includegraphics[scale=0.25]{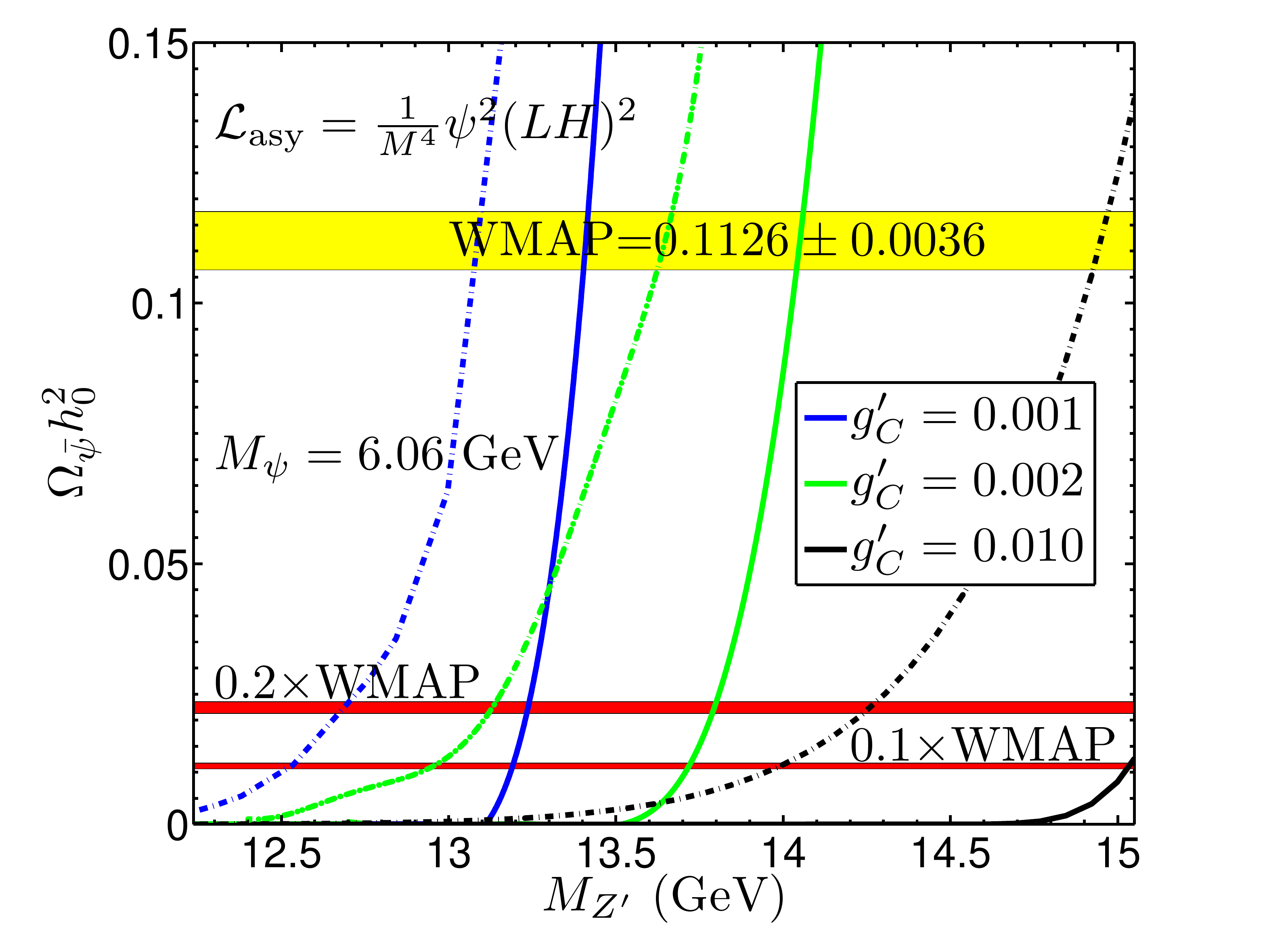}
\caption{\label{fig_bl_rel}
(color online)
A display of the thermal relic density of $\psi$ as a function of $M_{\zp}$ for the model
with  a gauged $B-L$ for different couplings with $\gamma=0$~(dashed line) and $\gamma=\gamma_0=1.3\times10^{-10}$~(solid line). It is seen that resonant annihilation of thermal dark matter
via the $\zp$ pole allows the relic density of this component to be reduced to below $10\%$ of the
WMAP value for values of $\zp$ around twice the mass of the dark particle.
}
\end{center}
\end{figure}

\section{Conclusion} \label{Sec:Con}
In this work we have proposed models of asymmetric dark matter in the framework of \st extensions
of the SM, the 2HD and the MSSM.
Several candidate models for asymmetric dark matter were discussed using a variety of operators
constructed out of standard model fields which carry a non-vanishing $B-L$ quantum number
which is transferred to the dark matter sector at high temperatures in the early universe consistent
with sphaleron interactions which preserve $B-L$. The analysis was done both for models
where the interaction temperature at which the $B-L$ transfer takes place above the electroweak phase
transition scale as well as below this scale. The details of the $B-L$ transfer determine the
mass of the asymmetric dark particle. A master formula was given which generates
the asymmetric dark matter mass for a variety of models discussed in the text and allows
one to discuss new possibilities. Specific models are discussed, including those anchored in
the standard model, the two Higgs doublet model as well as the minimal supersymmetric standard
model, with or without the right-handed neutrinos.
\\

A central ingredient in a successful asymmetric dark matter model,
and an explanation of cosmic coincidence, is an exhibition of an efficient mechanism for the
annihilation of the symmetric dark matter component which is produced by thermal processes.
We accomplish this using a \st extension of the \sm and of the MSSM.
The \st extension of the SM is particularly simple and appealing in which, aside from the
dark matter field, there is just one more field, a gauge boson ($Z'$) of a $L_{\mu}- L_{\tau}$ gauge
symmetry, which gains mass via the \st mechanism. The symmetric dark matter produced by
thermal processes is depleted via resonant annihilation from the exchange of a $Z'$
using a Breit-Wigner pole. This is perhaps the simplest asymmetric dark matter model, in that
there are no extra Higgs fields that appear in the model.
Moreover, the extra $U(1)$ gauge symmetry in our models forbids the dangerous
Majorana mass terms which would generate oscillations of the dark particles
and their anti-particles which could washout the asymmetry.
\\

A supersymmetric extension of this
model is also given where it is shown that in addition to the $Z'$ boson, there is also a
spin 0 boson field along with two additional neutralino states arising from the \st gaugino sector.
It is shown that the $\rho$ has a rapid decay and does not participate in the dark matter
analysis. In the MSSM extension, there is an extra complication, in that, with R parity one can
have a stable neutralino which is a possible dark matter candidate, and it must be shown
that it is also depleted so does not compete with the asymmetric dark matter candidate.
In the analysis presented in this work it was shown that
there exists a significant part of the parameter space of mSUGRA where the
relic density arising from the neutralino was less than one tenth of the WMAP relic density
and thus the neutralino is a subdominant component of dark matter. Interestingly, however, it was
shown that the subdominant neutralino is still accessible at future direct detection experiments
such as SuperCDMS and XENON-100.
It was shown that definitive tests of the model can come from a muon collider where one can produce
the $Z'$ which decays only into
$\mu$'s and $\tau$'s (and $\mu,\tau$ neutrinos).
\\

We also discussed a gauged $B-L$ \st model.
Again in this model the symmetric dark matter can be efficiently depleted by annihilation
near a Breit-Wigner $Z'$ pole.  Thus within the \st extensions there exist several possibilities for
explaining cosmic coincidence.
The dominant dark matter in all these models will be light
and lies in the range of $1-16\GeV$.
\\
\\

\noindent {\it Acknowledgements}:
PN acknowledges discussions with  Rabindra Mohapatra and Apostolos Pilaftsis.
WZF is grateful to HaiPeng An, Ning Chen, Hao Zhang and Peng Zhou for very helpful discussions.
This research is supported in part by the U.S. National Science Foundation (NSF) grants
PHY-0757959 and PHY-0969739  and through XSEDE under grant numbers TG-PHY110015.

\section*{Appendix}

\appendix

\section{Master formula for computing the asymmetric dark matter mass}\label{master}
We have discussed various Models A-F and subcases such as $\rm A_1$-$\rm A_6$ etc,
and also models with right-handed Dirac neutrinos, which will be discussed at the end of this section.
We discuss  now a master formula which allows one to take some particles in or out of thermal
equilibrium. Such a formula would generally be useful  before $SU(2)_{L}$  breaking, i.e., $T>T_{{\rm EWPT}}$,
when some of the super-particles are suppressed in the plasma while others are not.
This would allow us to discuss the Models A,D,E,F in a unified way and also allow us to generate
new models where some other sets  of super-particles are taken out of the relativistic plasma in the
early Universe. However, it is not useful to discuss such a formula below the electroweak
phase transition scale since the current experimental data indicates the sparticles to be heavy and
not below the electroweak phase transition scale.
\\

In obtaining the master formula, we assume: (1) In supersymmetric
cases, all particles in a supermultiplet have the same chemical potential;
(2) A given particle type in different
generations has the same chemical potential, e.g., $\mu_{d}=\mu_{s}=\mu_{b}$;
(3) All the additional Higgs doublets have the same chemical potential as the \sm Higgs $\mu_{H}$.
Following the discussion of Section~\ref{T>EWPT}, for all the fields
in the plasma, we have the chemical potential constraints as before, i.e,
$\mu_{H}=\mu_{L}-\mu_{e}=\mu_{q}-\mu_{d}=\mu_{u}-\mu_{q}$ from Yukawa couplings,
$3\mu_{q}+\mu_{L}=0$ from sphaleron processes,  and $Y=0$ from the
neutrality condition.
\\

It is useful to introduce the temperature-dependent coefficients $c_{\alpha}^{(i)}$
for the matter fields in the plasma. We define
$c_{\alpha}^{(i)}=c_{\alpha}^{(i)}(f)+c_{{\alpha}}^{(i)}(b)$,
where $c_{\alpha}^{(i)}(f)$ counts the contribution of $i^{\rm th}$ generation particle $\alpha$ (with
mass $m_{\alpha}$) which is fermionic and
$c_{\alpha}^{(i)}(b)$ counts the contribution of its super-partner $\tilde{\alpha}$
(with mass $m_{\tilde{\alpha}}$) which is bosonic, where
$c_{\alpha}(f)$ and  $c_{\alpha}(b)$ are given by Eq.~(\ref{cdef}).
We note that in the limit when $m_{\alpha}$ can be neglected,
one has a weakly interacting plasma so that $c_{\alpha}(f)=1$ and $c_{\alpha}(b)=2$.
Thus we have, for $T \gg m_{\tilde{\alpha}}$, $c_{\alpha}^{(i)}=1+2=3$;
for $m_{\tilde{\alpha}} \gg T \gg m_{\alpha}$, $c_{\alpha}^{(i)}=1+0=1$;
and for $T \ll m_{\alpha}$, $c_{\alpha}^{(i)}=0$.
For the Higgs doublets, we have $c_{H}=c_H (b)=2$
in the non-supersymmetric case, and $c_{H}=c_H (b)+c_H (f)=3$ for the supersymmetric
case,  and $\lambda_{H}$  counts the number of
Higgs doublets in the model.
\\

We can then rewrite the hypercharge neutrality condition as
\begin{align}
2(c_{q}^{(1)}+c_{q}^{(2)}+c_{q}^{(3)})\mu_{q}+4(c_{u}^{(1)}+c_{u}^{(2)}+c_{u}^{(3)})\mu_{u}-2(c_{d}^{(1)}+c_{d}^{(2)}+c_{d}^{(3)})\mu_{d} & \nonumber\\
-2(c_{L}^{(1)}+c_{L}^{(2)}+c_{L}^{(3)})\mu_{L}-2(c_{e}^{(1)}+c_{e}^{(2)}+c_{e}^{(3)})\mu_{e}+2c_{H}\lambda_{H}\mu_{H} & =0\,.
\end{align}
Further defining
$C_\alpha = \sum_i c^{(i)}_\alpha = c_{\alpha}^{(1)}+c_{\alpha}^{(2)}+c_{\alpha}^{(3)}$,
and together with Eqs.~\eqref{OPTcon1}-\eqref{OPTcon3}, we obtain
\begin{equation}
\mu_{X}=-\frac{C_q+8C_u+2C_d+3C_L+6C_e+3c_{H}\lambda_{H}}{6C_u+3C_d+3C_e+3c_{H}\lambda_{H}}Q_{B-L}^{{\rm DM}}\mu_{L}\,,
\end{equation}
and
\begin{align}
B-L & =-\Big[C_{u}\big(3C_{q}+6C_{d}+9C_{L}+2C_{e}\big)+C_{d}\big(3C_{q}+9C_{L}+8C_{e}\big)+C_{e}\big(C_{q}+3C_{L}\big)\nonumber \\
 & \quad+c_{H}\lambda_{H}\big(2C_{q}+C_{u}+C_{d}+6C_{L}+3C_{e}\big)\Big]\mu_{L}\Big/\big(6C_{u}+3C_{d}+3C_{e}+3c_{H}\lambda_{H}\big)\nonumber \\
 & \equiv - \; \mathcal{N}\,\mu_L \Big/\big(6C_{u}+3C_{d}+3C_{e}+3c_{H}\lambda_{H}\big)\,.
\end{align}
From Eqs.~\eqref{DMmass} and \eqref{B/B-L}, we find the master formula for computing
the dark matter mass
\begin{equation}
m_{{\rm DM}}
\simeq \frac{\mathcal{N}}{C_q+8C_u+2C_d+3C_L+6C_e+3c_{H}\lambda_{H}}
\cdot\frac{\kappa}{-Q_{B-L}^{\mathcal{O}}}\cdot\frac{150}{97}~{\rm GeV}\,,\label{Master}
\end{equation}
where $\kappa$ is the parameter indicating the dark matter type:
for the non-supersymmetric case, $\kappa=1$ for fermionic dark matter,
and $\kappa=2$ for bosonic dark matter; for the supersymmetric case, $\kappa=3$.
$Q_{B-L}^{\mathcal{O}}$ is the $(B-L)$-charge of the operator $\mathcal{O}_{\rm asy}$
 in the  $B-L$ transfer interaction Eq.~\eqref{ASYINT}.
The operators that carry $Q_{B-L}^{\mathcal{O}}= -1$ are: $\mathcal{O}_{\rm asy} =LH, LLe^c, Lqd^c, u^c d^c d^c$.
Higher dimensional operators with a larger value of $(B-L)$-charge can be constructed from them.
\\

We can extract the results for Models A,D,E,F from this master formula:
\begin{enumerate}
\item Model A:  For matter fields we have $c_{\alpha}^{(i)}=c_{\alpha}^{(i)}(f)=1$ (so that $C_\alpha=3$); $c_{H}=c_{H}(b)=2$ and $\lambda_{H}=1$. For fermionic dark matter we take $\kappa=1$ and we recover Eq.~\eqref{DMMFA}.
\item Model D:  Here all the parameters are the same as in Model A except that $\lambda_{H}=2$. Setting $\kappa=1$ we recover Eq.~\eqref{DMMFD}.
\item Model E:  For matter fields, $c_{\alpha}^{(i)}=c_{\alpha}^{(i)}(f)+c_{\alpha}^{(i)}(b)=3$ (so that $C_\alpha=9$); $c_{H}=c_{H}(b)+c_{H}(f)=3$ and $\lambda_{H}=2$. Since this is a supersymmetric case, $\kappa=3$ and we recover Eq.~\eqref{DMMFEF}.
\item Model F:  For matter fields, $c_{\alpha}^{(1)}=c_{\alpha}^{(2)}=1, c_{\alpha}^{(3)}=3$ (so that $C_\alpha=5$); $c_{H}=3$, $\lambda_{H}=2$. Taking $\kappa=3$ we recover Eq.~\eqref{DMMFEF}.
\end{enumerate}
\vspace{0.5cm}

For Models $\rm A',D',E',F'$ which include three generations of
right-handed Dirac neutrinos in the thermal bath,
the master formula reads,
\begin{equation}
m_{{\rm DM}}
\simeq \left(\frac{\mathcal{N}}{C_q+8C_u+2C_d+3C_L+6C_e+3c_{H}\lambda_{H}}+C_{\nu_R}\right)
\cdot\frac{\kappa}{-Q_{B-L}^{\mathcal{O}}}\cdot\frac{25}{21}~{\rm GeV}\,,\label{Master'}
\end{equation}
where $C_{\nu_R}=c_{\nu_R}^{(1)}+c_{\nu_R}^{(2)}+c_{\nu_R}^{(3)}$.
By taking $C_{\nu_R}=3,3,9,5$, we can recover the dark matter mass formulas
for Models $\rm A',D',E',F'$ (Eqs.~\eqref{DMMFAD'} and \eqref{DMMFEF'}).
\\

We can also obtain the dark matter mass of other models
from the master formula Eq.~\eqref{Master} or Eq.~\eqref{Master'}
by varying the temperature where certain
heavy particles are Boltzmann suppressed in the thermal bath.

\section{Decay of the  $\rho$} \label{App:rhoDR}
Here we compute the decay of the $\rho$.
From Eq.~(\ref{4.k}) one finds that $\rho$ couples
to smuons, staus, muon sneutrino, and tau sneutrino. This means that the $\rho$ decay has
$\mu^+\mu^-, \nu_{\mu}\bar\nu_{\mu},\, \tau^+\tau^-, \nu_{\tau} \bar \nu_{\tau}$ final states which arise
via the exchange of neutralinos and charginos in the loops (a generic diagram is shown in
Fig.~\ref{FD_rhodecayP}).
The amplitude of the generic diagram reads,
\begin{equation}
i\mathcal{M}=-ig_{\rho ij}C_{ki}C^*_{kj}\int\frac{d^{4}k}{(2\pi)^{4}}\bar{u}(p')\frac{(\slashed k-\slashed p)+m_{\tilde \chi_k}}{(k^{2}-m_i^{2})(k'^{2}-m_j^{2})((k-p)^{2}-m_{\tilde \chi_k}^{2})}v(p)\,,
\end{equation}
where $k'=q-k$,  $m_i, m_j$ are the masses of the sleptons, and $m_{\tilde \chi_k}$ is the mass of the
neutralino or of the chargino in the loop, while $g_{\rho ij},C_{ki}$ are the couplings.  Our purpose here is to
estimate the size of the lifetime and to that end it is sufficient to estimate the contribution for one set of
diagrams.  Thus we consider the decay of the $\rho$ to  final states $\mu^+\mu^-$ via the exchange
of neutralinos. In this case we will have the exchange of smuons and neutralinos in the loop.
Further, we will ignore the mixing between the left and the right chiral smuons so that the mixing angle
$\theta_{\tilde f_i}$ in Eq.~(\ref{4.k})  can be set to zero. In this circumstance the off-diagonal term
involving two smuons in the loop does not contribute and the relevant loop integral takes the form

\begin{figure}[t!]
\begin{center}
\includegraphics[scale=0.44]{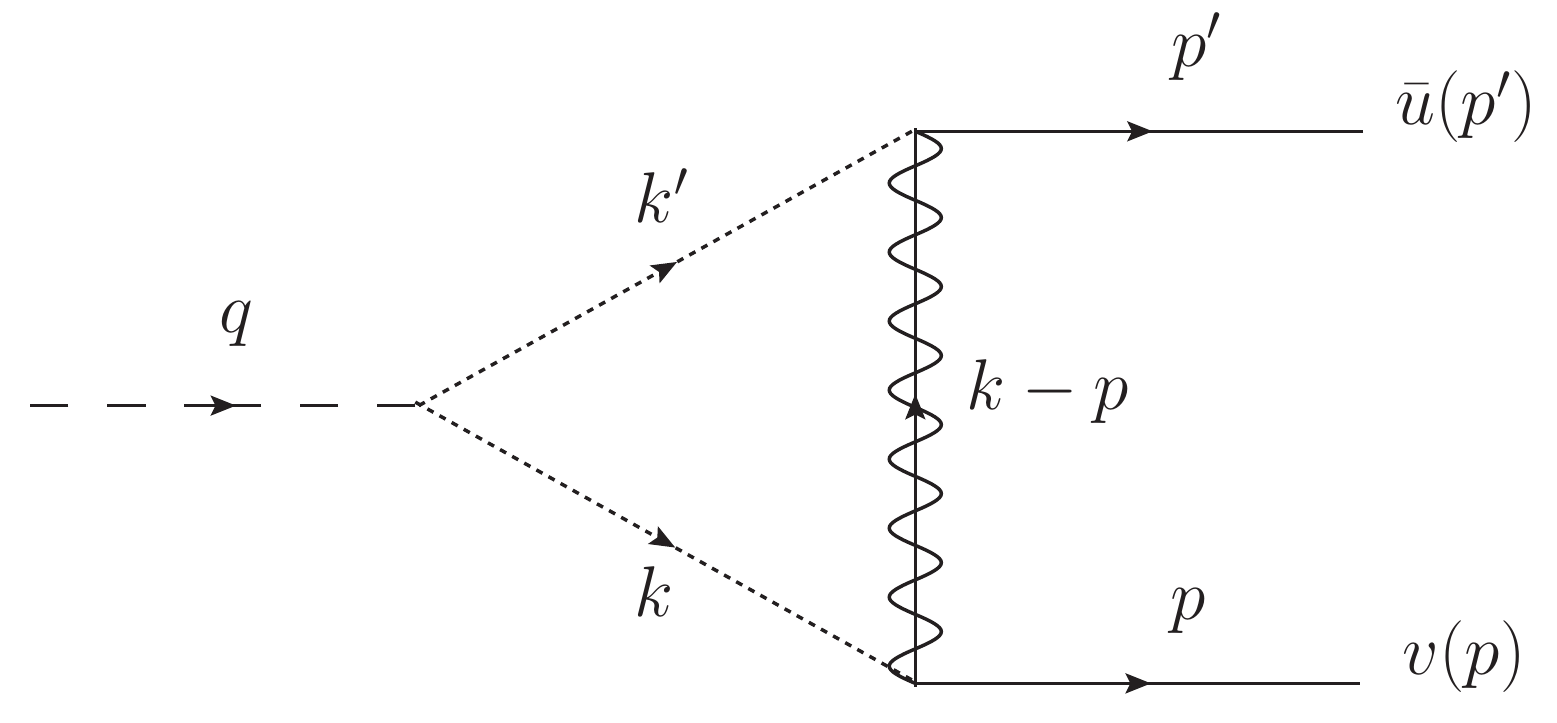}
\caption{\label{FD_rhodecayP}
A generic diagram showing the decay of the
$\rho$  to one of the final states which could be $\mu^+\mu^-, \nu_{\mu} \bar\nu_{\mu},\,
\tau^+\tau^-, \nu_{\tau} \bar \nu_{\tau}$ via exchange of sleptons, charginos and neutralinos at one loop.}
\end{center}
\end{figure}

\begin{equation}
\frac{1}{(k^{2}-m_i^{2})(k'^{2}-m_j^{2})((k-p)^{2}-m^2_{{\tilde \chi}_k})}=\int_{0}^{1}{\rm d}x{\rm d}y{\rm d}z\delta(x+y+z-1)\frac{2}{D_{ik}^{3}}\,,
\end{equation}
where $D_{ik}=l^{2}-\Delta_{ik}+i\epsilon$ in which $l\equiv k-(yq+zp)$ and
\begin{equation}
\Delta_{ik}=(1-z)m_i^{2}-xym_{\rho}^{2}+zm_{\tilde \chi_k}^{2}+(z^{2}-z)m_{\mu}^{2}\,.
\end{equation}
The masses in the loops are much larger than the muon mass and thus the muon mass
can be ignored.
The integration on $l$ gives
\begin{align}
i\mathcal{M} & =\frac{-ig_{\rho ij}C_{k i}C^*_{k i}}{(4\pi)^{2}}\int_{0}^{1}{\rm d}x{\rm d}y{\rm d}z\delta(x+y+z-1)\frac{\bar{u}(p')m_{\tilde \chi_k}v(p)}{\Delta_{ik}}\,.
\end{align}
Further, an approximate evaluation of
integration on the Feynman parameters gives
\begin{align}
i\mathcal{M} = \frac{-ig_{\rho ii}C_{k i}C^*_{ki}}{(4\pi)^{2}}\bar{u}(p')\frac{m_{\tilde \chi_k}}{m_i^2}v(p)\,,
\end{align}
under the assumption $m^2_{\tilde \chi_k}\big/m_i^2 \ll 1$ and $m^2_{\rho}\big/m_i^2 \ll 1$.
The decay width of $\rho \to \mu^+\mu^-$ is then given by
\begin{align}
{\rm d}\Gamma & =\frac{1}{2m_{\rho}}\int\frac{{\rm d}^{3}\vec{p}}{(2\pi)^{3}2E_{\mu^{+}}}\int\frac{{\rm d}^{3}\vec{p'}}{(2\pi)^{3}2E_{\mu^{-}}}\left|\sum i\mathcal{M}\right|^{2}(2\pi)^{4}\delta^{(4)}(q-p-p')
  =\frac{\left|\sum i\mathcal{M}\right|^{2}}{8\pi m_{\rho}}\,.
\label{rhodecayR}
\end{align}
Next we note that $g_{\rho 11}= -g_{\rho 22}= g_C Q_C m_{\rho}$ and thus
\begin{equation}
\left|\sum i\mathcal{M}\right|^{2} \simeq
\frac{(g_C Q_c)^2 m^4_\rho}{16 \pi^4} \left|\sum_{k=1}^{6} \sum_{i=1}^{2} (-1)^{i+1} \frac{C_{k i} C^*_{k i}}{m_i^2}\right|^2\,.
\label{SqAmp}
\end{equation}
A numerical estimate using Eqs.~(\ref{rhodecayR}) and (\ref{SqAmp}) and the inputs  $m_1 = 1 $ TeV, $m_2 \gg m_1$,
$m_{\rho} = 100$~GeV, the lightest neutralino mass of 50~GeV
gives
$\tau_\rho = {\hbar}/{\Gamma} \sim 10^{-14\pm 1}~{\rm s}.$
Thus the decay of the $\rho$ is very rapid.

\section{$Z'$ exchange contribution to $\mu^+\mu^-\to e^+e^-$ at loop level} \label{App:Z'}

At a muon collider, $e^{+}e^{-}$ final states can be created  via photon exchange and via a $Z$ boson
exchange. Since the $Z'$ has no direct coupling with the first generation leptons,  there is no
tree-level $Z'$ exchange contribution to $e^{+}e^{-}$ final states.
However, at the loop level a $Z'$ exchange can make a contribution where the second and third
generation leptons are exchanged in the loop as shown in Fig.~{\ref{Bfig}.
\begin{figure}[t!]
\begin{center}
\includegraphics[scale=0.67]{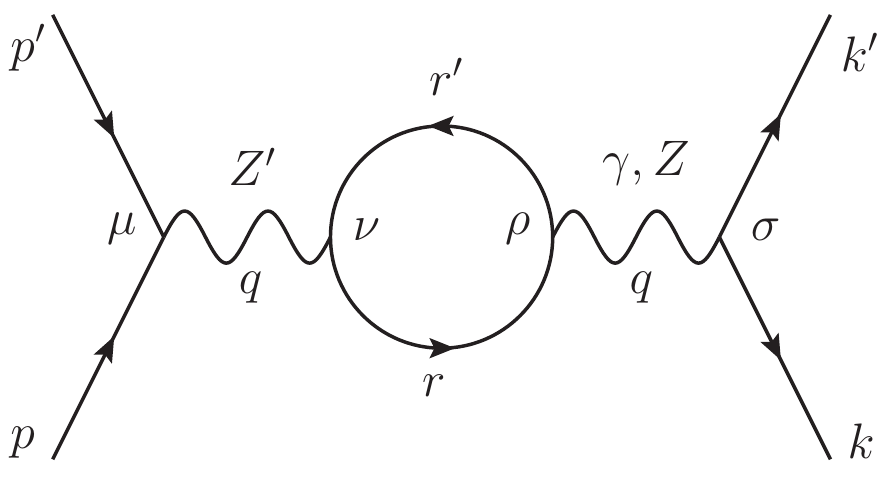}
\caption{\label{FD_ZprimeMIX}
$Z'-\gamma$ and $Z'-Z$ exchange via $\mu^+\mu^-, \nu_{\mu} \bar\nu_{\mu},\,
\tau^+\tau^-, \nu_{\tau} \bar \nu_{\tau}$ loops.}
\label{Bfig}
\end{center}
\end{figure}
We now compute this contribution to determine its size.
Thus we consider a $\mu^{+}\mu^{-}\rightarrow e^{+}e^{-}$ process with a $Z'$ exchange
via the second and third generation leptons loops as shown in Fig.~{\ref{Bfig}.
In this case  the contribution to the scattering amplitude is
\begin{align}
\sum i\mathcal{M} & =i\mathcal{M}_{\gamma Z
'}+i\mathcal{M}_{ZZ'}\nonumber\\
 & =\bar{v}(p')(\frac{i}{2}g_{C}Q_{\mu}\gamma^{\mu})u(p)\frac{-i(g_{\mu\nu}-q_{\mu}q_{\nu}/m_{Z'}^{2})}{q^{2}-m_{Z'}^{2}}(i\Pi_{\gamma Z
'}^{\nu\rho})\frac{-ig_{\rho\sigma}}{q^{2}}\bar{u}(k)(-ie\gamma^{\sigma})v(k')\nonumber\\
 & \,+\bar{v}(p')(\frac{i}{2}g_{C}Q_{\mu}\gamma^{\mu})u(p)\frac{-i(g_{\mu\nu}-q_{\mu}q_{\nu}/m_{Z'}^{2})}{q^{2}-m_{Z'}^{2}}(i\Pi_{ZZ'}^{\nu\rho})\nonumber\\
 & \qquad\qquad\qquad\qquad\qquad\quad\times\frac{-i(g_{\rho\sigma}-q_{\rho}q_{\sigma}/m_{Z}^{2})}{q^{2}-m_{Z}^{2}}\bar{u}(k)\frac{-ig\gamma^{\sigma}}{\cos\theta_{W}}(g_{V}+g_{A}\gamma^{5})v(k')\,,
\end{align}
where $Q_{\mu}$ is the $U(1)_{L_{\mu}-L_{\tau}}$ charge for muon,
$g_V=\tfrac{1}{2}(T_3)_L+\sin^2 \theta_W$, $g_A=-\tfrac{1}{2}(T_3)_L$, and the vacuum polarization tensors
$i\Pi_{\gamma Z
'}^{\nu\rho}$ and $i\Pi_{ZZ'}^{\nu\rho}$ are the sums of
the contributions from $\mu^{+}\mu^{-},\nu_{\mu}\bar{\nu}_{\mu},\,\tau^{+}\tau^{-},\nu_{\tau}\bar{\nu}_{\tau}$
loops
\begin{equation}
i\Pi^{\nu\rho}=i\Pi_{\mu}^{\nu\rho}+i\Pi_{\nu_{\mu}}^{\nu\rho}+i\Pi_{\tau}^{\nu\rho}+i\Pi_{\nu_{\tau}}^{\nu\rho}\,.
\end{equation}
First we focus on $i\Pi_{\gamma Z
',\mu}^{\nu\rho}$ which is the muon loop contribution
to the $Z'-\gamma$ exchange.  It reads
\begin{align}
i\Pi_{\gamma Z
',\mu}^{\nu\rho} & =-(\frac{i}{2}g_{C}Q_{\mu})(-ie)\int\frac{{\rm d}^{4}r}{(2\pi)^{4}}{\rm tr}\Big[\gamma^{\nu}\frac{i(\slashed r+m_{\mu})}{(r^{2}-m_{\mu}^{2})}\gamma^{\rho}\frac{i(\slashed r'+m_{\mu})}{(r'^{2}-m_{\mu}^{2})}\Big]\nonumber \\
 & =\frac{4ieg_{C}Q_{\mu}}{(4\pi)^{2}}(q^{2}g^{\nu\rho}-q^{\nu}q^{\rho})\int_{0}^{1}{\rm d}x\, x(1-x)\frac{{\rm \Gamma}(2-\frac{d}{2})}{\Delta_{\mu}^{2-\frac{d}{2}}}\nonumber \\
 & =\frac{4ieg_{C}Q_{\mu}}{(4\pi)^{2}}(q^{2}g^{\nu\rho}-q^{\nu}q^{\rho})\int_{0}^{1}{\rm d}x\, x(1-x)\big(\frac{2}{\epsilon}-{\rm log}\Delta_{\mu}-\gamma+{\rm log}(4\pi)+\mathcal{O}(\epsilon)\big)\,,
\end{align}
where $\Delta_{\mu}=m_{\mu}^{2}-x(1-x)q^{2}$, and in the last step
we use the dimensional regularization. The expression of $i\Sigma_{\gamma Z
',\tau}^{\nu\rho}$
differs from $i\Sigma_{\gamma Z
',\mu}^{\nu\rho}$ by only the $Q_{\tau}$ factor,
and it takes the form
\begin{equation}
i\Pi_{\gamma Z
',\tau}^{\nu\rho}=\frac{4ieg_{C}Q_{\tau}}{(4\pi)^{2}}(q^{2}g^{\nu\rho}-q^{\nu}q^{\rho})\int_{0}^{1}{\rm d}x\, x(1-x)\big(\frac{2}{\epsilon}-{\rm log}\Delta_{\mu}-\gamma+{\rm log}(4\pi)+\mathcal{O}(\epsilon)\big)\,.
\end{equation}
Summing over these two terms, we find  a dramatic cancellation of the  divergence in the loop due to
$Q_{\mu}=-Q_{\tau}=1$, making the loop finite so that
\begin{equation}
i\Pi_{\gamma Z
',\mu}^{\nu\rho}+i\Pi_{\gamma Z
',\tau}^{\nu\rho}=\frac{4ieg_{C}}{(4\pi)^{2}}(q^{2}g^{\nu\rho}-q^{\nu}q^{\rho})\times I\,,
\end{equation}
where
\begin{equation}
I=\int_{0}^{1}{\rm d}x\, x(1-x){\rm log}\frac{\Delta_{\tau}}{\Delta_{\mu}}=\int_{0}^{1}{\rm d}x\, x(1-x){\rm log}\frac{m_{\tau}^{2}-x(1-x)q^{2}}{m_{\mu}^2-x(1-x)q^{2}}\,.
\end{equation}
One can also obtain the neutrino exchange contributions from the above by setting the
fermion masses to zero in the equation above (assuming neutrinos to be massless) which
gives a vanishing contribution.
\\

Now we want to compare the contribution of the $Z'-\gamma$ exchange loop
diagram with the tree-level process $\mu^{+}\mu^{-}\rightarrow \gamma \rightarrow e^{+}e^{-}$,
whose amplitude reads
\begin{equation}
i\mathcal{M}_{\gamma}=\bar{v}(p')(-ie\gamma^{\mu})u(p)\frac{-ig_{\mu\nu}}{q^{2}}\bar{u}(k)(-ie\gamma^{\nu})v(k')\,.
\end{equation}
With some manipulation we find
\begin{equation}
i\mathcal{M}_{\gamma Z
'}=-\frac{2g_{c}^{2}I}{(4\pi)^{2}}\cdot\frac{q^{2}}{q^{2}-m_{Z'}^{2}}\times i\mathcal{M}_{\gamma}\equiv f\times i\mathcal{M}_{\gamma}\,.
\end{equation}
Thus, the total squared amplitudes involving a photon can be written
as
\begin{align}
|i\mathcal{M}_{\gamma}+i\mathcal{M}_{\gamma Z
'}|^{2} & =|1+f|^{2}\times|i\mathcal{M}_{\gamma}|^{2}\nonumber\\
 & =(1+f+f^{*}+ff^{*})\times|i\mathcal{M}_{\gamma}|^{2}\,.
\end{align}
Our numerical analysis shows that $(f+f^{*}+ff^{*})$ is smaller than
$\sim10^{-3}$ and thus the loop makes only a tiny contribution to the total cross
section in this case. The analysis of $Z'-Z$ exchange is similar and gives a very small value. Thus we conclude that a $Z'$ peak will not be visible in the
$\mu^+\mu^-\to e^+e^-$ process at a muon collider. The above analysis also exhibits
why a $\zp$ in this model would not be visible in an $e^+e^-$ machine.

\end{document}